\documentclass[conference,10pt]{IEEEtran}
\IEEEoverridecommandlockouts
\usepackage{cite}
\usepackage{amsmath,amssymb,amsfonts}
\usepackage{algorithmic}
\usepackage{xcolor}
\usepackage{algorithm}
\usepackage{array}
\usepackage[caption=false,font=normalsize,labelfont=sf,textfont=sf]{subfig}
\usepackage{textcomp}
\usepackage{stfloats}
\usepackage{color}
\usepackage{tabularx}
\usepackage{booktabs} 
\usepackage{url}
\usepackage{verbatim}
\usepackage[T1]{fontenc}
\usepackage{graphicx}
\usepackage{caption}
\usepackage{wrapfig}
\usepackage{makecell}
\usepackage{balance}
\usepackage{multirow}
\usepackage{subcaption}
\usepackage{amsfonts} % 导入amsfonts包以使用QED符号
\usepackage{amsmath}
\usepackage{hyperref}
\newcommand{\figref}[1]{Fig.~\ref{#1}}
\newcommand{\tabref}[1]{Tab.~\ref{#1}}
\newcommand{\equref}[1]{Eq.~(\ref{#1})}
\newcommand{\defref}[1]{Def.~\ref{#1}}
\newcommand{\theref}[1]{Theorem~\ref{#1}}
\newtheorem{Definition}{Definition}
\newtheorem{Theorem}{Theorem}

\def\BibTeX{{\rm B\kern-.05em{\sc i\kern-.025em b}\kern-.08em
    T\kern-.1667em\lower.7ex\hbox{E}\kern-.125emX}}
\begin{document}

\title{Bridging Quantum Computing and Differential Privacy: Insights into Quantum Computing Privacy}

\author{ 
\IEEEauthorblockN{Yusheng Zhao$^*$, Hui Zhong$^\dagger$, Xinyue Zhang$^\ddagger$, Yuqing Li$^*$, Chi Zhang$^*$, Miao Pan$^\dagger$
} \IEEEauthorblockA{
 \textit{$^*$University of Science and Technology of China}, Hefei,  P. R. China \\ 
 \textit{$^\dagger$University of Houston}, Houston, USA \\
 \textit{$^\ddagger$Kennesaw State University}, Marietta, USA} 
 }

\maketitle

\begin{abstract}
While quantum computing has strong potential in data-driven fields, the privacy issue of sensitive or valuable information involved in the quantum algorithm should be considered. Differential privacy (DP), which is a fundamental privacy tool widely used in the classical scenario, has been extended to the quantum domain, i.e., quantum differential privacy (QDP). QDP may become one of the most promising approaches toward privacy-preserving quantum computing since it is not only compatible with classical DP mechanisms but also achieves privacy protection by exploiting unavoidable quantum noise in noisy intermediate-scale quantum (NISQ) devices. This paper provides an overview of the various implementations of QDP and their performance in terms of privacy parameters under the DP setting. Specifically, we propose a taxonomy of QDP techniques, categorizing the literature on whether internal or external randomization is used as a source to achieve QDP and how these implementations are applied to each phase of the quantum algorithm. We also discuss challenges and future directions for QDP. By summarizing recent advancements, we hope to provide a comprehensive, up-to-date review for researchers venturing into this field.
\end{abstract}


\section{Introduction}
    Leveraging its key attributes, such as interference, superposition, and entanglement, \textit{quantum computing (QC)} has emerged as a technology with strong potential for exponential acceleration in addressing some particular problems, such as database search~\cite{grover1996fast}, prime factorization~\cite{shor1994algorithms}, and solving linear equations~\cite{harrow2009quantum}. In recent years, rapid progress has been made in verifying the quantum advantage of realistic quantum devices, e.g., faster Gaussian Boson sampling (GBS) on \textit{Jiuzhang} of the USTC~\cite{zhong2020quantum} and the proven quantum supremacy on \textit{Sycamore} of Google~\cite{arute2019quantum}. Despite these fruitful advancements, we are still in the so-called \textit{noisy intermediate-scale quantum (NISQ)} era. To adapt near-term NISQ computers, several practical quantum algorithms have been developed, with the most well-known being the \textit{variational quantum algorithm (VQA)}~\cite{cerezo2021variational}, including \textit{variational quantum eigensolver (VQE)}~\cite{peruzzo2014variational}, \textit{quantum approximate optimization algorithm (QAOA)}~\cite{farhi2014quantum} and \textit{quantum machine learning (QML)}~\cite{rebentrost2014quantum,lloyd2018quantum,cong2019quantum,bausch2020recurrent}.
    %which are used for basic energy estimation of computational chemistry and many-body physics, graph and combinatorial optimization problem, and potential speedup on machine learning model enabling quantum hardwares, respectively.
     
    With increasing interest in the role of quantum algorithms in data-driven fields, such as drug discovery~\cite{cao2018potential}, materials engineering~\cite{de2021materials}, financial analysis~\cite{orus2019quantum, herman2023quantum}, and climate prediction~\cite{berger2021quantum}, privacy issues have arisen. Training data and measurement-extracted information are at risk of leakage and security attacks~\cite{xu2023classification, chu2023qtrojan, chu2023qdoor}. Moreover, many current NISQ devices are already cloud-based~\cite{amazon2020braket,ms2023azure}, which increases the susceptibility to attacks by malicious third parties. Undoubtedly, the privacy issues in the quantum scenario demand serious consideration.

    Amidst growing concerns surrounding privacy in quantum algorithms, the exploration of effective safeguarding methods has become critical. \textit{Differential privacy (DP)}~\cite{dwork2014algorithmic} has become a beacon of hope as a well-practiced approach in the classical scenario. DP guarantees that subtle changes in an individual's information have little influence on the result of the algorithm. Under the DP setting, the definition and analysis of privacy are rigorous and mathematical, with privacy strength quantitatively represented by privacy parameters. This aspect has led DP to become a common approach for privacy regulation and preservation in various data-driven paradigms, such as databases~\cite{dwork2006calibrating} and deep learning~\cite{abadi2016deep}. 

    As DP has proven effective in classical settings, its quantum extension, \textit{quantum differential privacy (QDP)}~\cite{zhou2017differential}, is emerging as a promising approach toward privacy-preserving quantum computing. Especially in the NISQ era, while unavoidable quantum noise limits the capability of quantum algorithms and poses a computational difficulty, it may be a natural source for achieving QDP~\cite{du2021quantum}. In this work, a quantum algorithm $\mathcal{A}$ with quantum noise (normally in the circuit) can be described by a triplet $(\phi, \mathcal{E}_{\mathcal{N}},\{M_i\}_{i\in\mathcal{O}})$, orderly composed of a quantum encoding $\phi$ with (optional) classical input for state preparation, a noisy quantum circuit $\mathcal{E}_{\mathcal{N}}$ for state transformation, and a set of measurements $\{M_i\}_{i\in\mathcal{O}}$ for information extraction. For neighboring quantum states $\rho$ and $\sigma$ within distance $d$ generated by $\phi$, we say that the quantum algorithm $\mathcal{A}$ satisfies $(\epsilon,\delta)$-QDP only and if only:
    \begin{equation*}
            \sum_{i\in \mathcal{O}}\mathrm{Tr}[M_i\mathcal{E}(\rho)] \leq \exp{(\epsilon(d))} \cdot \sum_{i\in \mathcal{O}}\mathrm{Tr}[M_i\mathcal{E}(\sigma)] + \delta. 
    \end{equation*}
    See Section~\ref{sec:qdp} for the full details. 
    However, with the rise of the research topic of privacy-preserving quantum computing, researchers from diverse backgrounds (mainly physics and computer science) have published numerous papers on QDP. The intersectionality of this topic has illuminated different perspectives on DP-preserving quantum algorithms.

    Given this consideration, we have conceived this survey. To the best of our knowledge, this is the first survey on the topic of QDP, and we provide comprehensive insights into the topic at hand to researchers in related fields. We answer the following research questions (RQs), which are further described below.
    \begin{itemize}
        \item \textbf{RQ1}: What is QDP and how can it be formulated mathematically?
        \item \textbf{RQ2}: How can QDP be instantiated via internal or external randomization of the quantum algorithm?
        \item \textbf{RQ3}: How does QDP preserve privacy in quantum algorithms, and what is the performance of the privacy parameters in these QDP publications?
    \end{itemize}
     In \textbf{RQ1}, we give a detailed formalization of QDP (Section~\ref{subsec:qdp}). Concerning \textbf{RQ2}, we first introduce randomization mechanisms for classical and quantum scenarios (Section~\ref{subsec:formal_cdp} \&~\ref{subsec:quantum_noise}) and then present a case study on how to achieve QDP via a depolarizing channel (Section~\ref{subsec:case_study}). Regarding \textbf{RQ3}, we discuss some brief deductions of the privacy bounds of the privacy parameters of QDP induced by quantum noise (Section~\ref{subsec:qc_dp}) and consequently make a comparison in~\tabref{tab:qdp_qc}.

    % organization
    \textbf{Organization of this paper.} In Section \ref{sec:fundamental}, we first review some notions of quantum theory, including the basic concepts of quantum computation and quantum algorithm. We then introduce the formulation of classical DP. In Section~\ref{sec:qdp}, the quantum counterpart of classical DP is presented, and an additional case study on how to obtain QDP via quantum noise is given. In Section~\ref{Sec:DP_QA}, we categorize the literature and report how QDP plays a role in phases of the quantum algorithm. In addition, challenges and future potential directions are provided in Section~\ref{sec:challenges} before we conclude in Section~\ref{sec:conclusion}.

\section{Background}
\label{sec:fundamental}
    In this section, we first introduce some basic concepts of quantum computation and elaborate on how a quantum algorithm works. We then briefly formalize the definition of classical differential privacy. 
    \begin{figure*}
        \centering
        \includegraphics[width=1\linewidth]{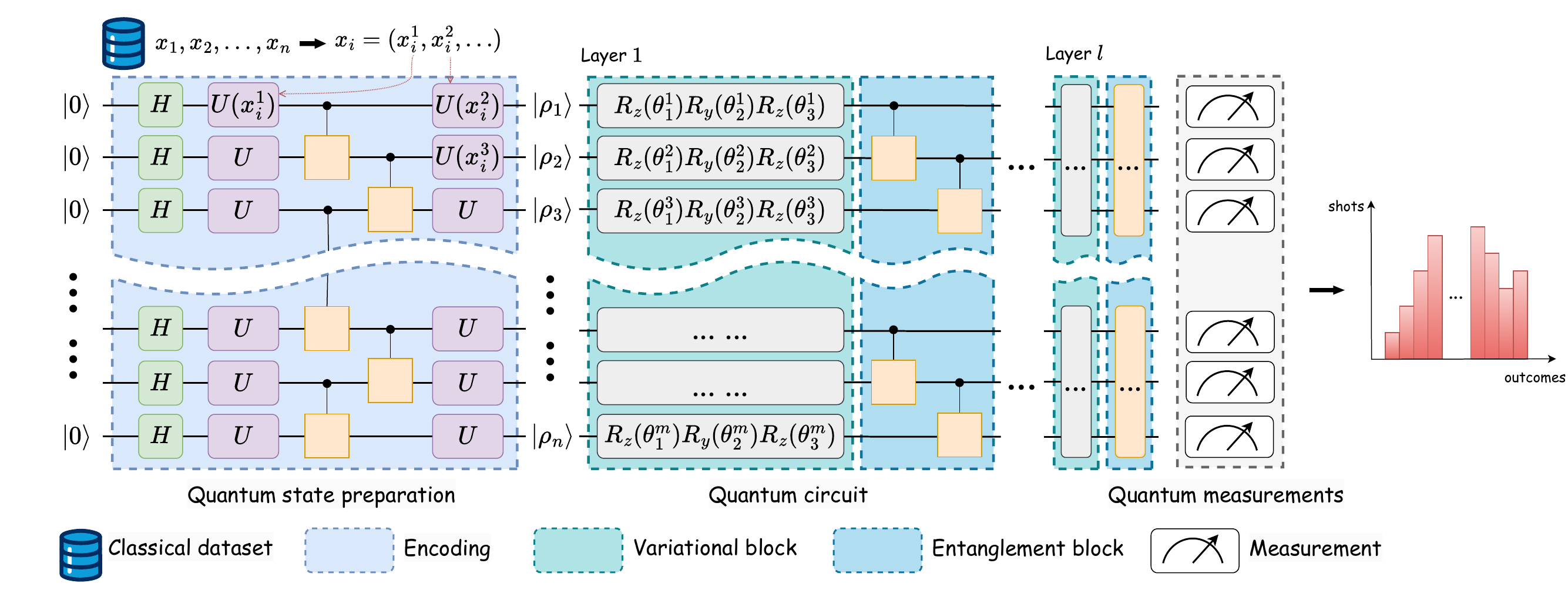}
        \caption{Variational quantum algorithm (VQA), taking QML model as an example. The classical part is omitted.}
        \label{fig:VQA}
    \end{figure*}   

    \subsection{Qubits, quantum gates and circuits}
        The atom unit of quantum computing is the quantum bits (\textit{qubits}). Qubits can be described as the linear combination of a set of computational basis state vectors (e.g., $|0\rangle$ and $|1\rangle$) in $2^n$-dimensional Hilbert space $\mathcal{H}$, where $n$ is the number of qubits. Quantum computing is an algorithmic process of quantum states characterized by qubits and is implemented on quantum hardware through quantum circuits consisting of wires and quantum gates.

        Quantum gates are mathematically represented by unitary matrices $U$. The gates satisfy $U^\dagger = U^{-1}$, where $U^\dagger$ is the adjoint (conjugate transpose) of $U$. The most basic quantum gates are \textit{1-qubit gates}, such as the Pauli gate ($I$, $X$, $Y$, $Z$), the Hadamard gate ($H$), and the rotation gate ($R_x$, $R_y$, $R_z$). Among these, the rotation gate is usually used for parameterized quantum circuits with learnable rotation angles $\theta$. More complex quantum gates are \textit{2-qubit controlled-U gates}, which induce entanglement between qubits, such as $CX$, $CZ$, and $CR_x$. More details can be found in the textbook~\cite{nielsen2010quantum}.

        A quantum circuit is a set of ordered collections of wires and quantum gates, and is utilized to evolve quantum states. For an input $n$-qubit state $\rho$ and a quantum circuit $\mathcal{E}=U_d\dots U_1$, where $U_i$ is the $i$-th quantum gate and $d$ is the depth of circuit $\mathcal{E}$, the output of evolution is represented as $\mathcal{E}(\rho)=U_d(\dots (U_1\rho U_1^\dagger) \dots)U_d^\dagger$. 
        
    \subsection{Quantum algorithm}
    \label{subsec:qa}
        A quantum algorithm is a set of tailored instructions for quantum computing that runs on quantum hardware, and aims at exponential acceleration to solve some particular problems. VQA~\cite{cerezo2021variational}, as a common type of practical quantum algorithm, is composed of initial quantum states, a parameterized quantum circuit (PQC), and quantum measurements. As illustrated in~\figref{fig:VQA}, the PQC is fed with quantum data or classical data encoded by quantum encoding, and its outputs are collapsed into classical information via quantum measurements.
       
        \paragraph{State preparation}
            Classical datasets are readily available and well-established but are not directly suitable for performing quantum algorithms such as VQAs. Thus it typically requires state preparation, i.e., initializing classical datasets into state vectors in Hilbert space via quantum encoding. Let $X = \{x_1,\dots, x_l\}$ be a classical dataset where $x_i$ is the $i$-th entry with $n$-dimensional features. Let $\phi$ be a quantum encoding method, initial states $\rho$ can be realized via
            \begin{equation}
            \label{eq:state_prep}
                \rho \leftarrow \phi(x)=|\phi(x)\rangle \langle\phi(x)|.
            \end{equation}
            Several common quantum encodings are summarized in~\tabref{tab:quencoding}.
            \begin{table}  
                \centering
                \caption{Common approaches for quantum encoding. The encoding method is denoted by $\phi$ in the rest of the paper.}
                \vspace{-3pt}
                \footnotesize
                \label{tab:quencoding}
                \begin{tabular}{c|c|c}
                    \toprule
                    Encoding & Mapping $\phi$  & Kernel $\kappa$  \\
                    \midrule
                    Basic encoding & $x\mapsto\frac{1}{\sqrt{n}}\sum^n_{k=1}|x^k\rangle$ & $\delta_{x_i,x_j}$  \\   
                    Amplitude encoding & $x\mapsto\sum^n_{k=1}\frac{x^k}{\parallel x \parallel}|k\rangle$ & $|x_i^\top x_j|^2$  \\
                    Angle encoding & $x\mapsto\bigotimes^n_{k=1}R(x^k)|0\rangle$ & $\prod^n_{k=1}|\cos(x_i^k-x_j^k)|^2$   \\
                    \bottomrule
                \end{tabular}
            \end{table}
            
             Moreover, similar to kernel methods, the notion of \textit{quantum kernel} was proposed in~\cite[Definition 2]{schuld2021supervised}. Let $x,y\in \mathcal{D}$, where $\mathcal{D}$ is the space of input classical data. The quantum kernel $\kappa$ is represented by the inner product between encoding vectors of $x$ and $y$:
             \begin{equation}
             \label{eq:kernel}
                 \kappa(x,y)=|\langle \phi(x) | \phi(y) \rangle|^2. 
             \end{equation}
            
        \paragraph{Quantum circuit}

            As previously described, the quantum circuit is the core computational component of the quantum algorithm. The input of the circuit is usually a pure state, which is a complex unit vector in $2^n$-dimensional Hilbert space $\mathcal{H}$, written in Dirac notation as $|\rho\rangle=\sum_i^n\alpha_i|i\rangle$, where $|i\rangle$ is an orthonormal basis in $\mathcal{H}$ and $\sum_i \alpha_i = 1$. For example, for a 1-qubit quantum system, a pure state in 2-dimensional Hilbert space can be represented as 
                $|\rho\rangle = \alpha|0\rangle + \beta |1\rangle$, where $|0\rangle=[1,0]^\top,|1\rangle=[0,1]^\top$ and $\alpha^2+\beta^2=1$. The circuit performed on naturally noisy hardware turns a pure state into a \textit{mixed state} as output, which is mathematically modeled as a $n\times n$-dimensional density matrix $\psi=\sum_k p_k|\rho_k\rangle\langle\rho_k|$, where $p_k$ is the probability that $\psi$ collapses into $|\rho_k\rangle$ and $\sum_k p_k^2=1$.

        \paragraph{Quantum measurements}
            The output of the quantum algorithm is determined by the probabilistic occurrence of quantum states, resulting from measurements over superposition. Therefore, the final answer to the algorithmic flow is a discrete probabilistic distribution, as illustrated in the last part in \figref{fig:VQA}. The x-axis represents all the possible outcomes of the measurements and the y-axis represents the number of shots. Formally, if the output state of a circuit is $\rho$ and a set of measurements are represented by $n \times n$ positive semi-definite matrices $\{M_i\}_{i\in \mathcal{O}}$, where $\mathcal{O}$ is the set of all possible outcomes, then $p=\{p_i\}_{i\in\mathcal{O}}$ is the result of the measurements, where $p_i=\mathrm{Tr}(M_i\rho)$ is the probability that the outcome of the measurement is $i$.

    \subsection{The formulation of classical DP}
    \label{subsec:formal_cdp}
        Differential privacy (DP)~\cite{dwork2006calibrating} has been a de facto standard for evaluating the privacy strength of an algorithm. DP provides a quantifiable privacy guarantee by ensuring that subtle changes in datasets do not affect the probability of any outcome. We first review some basic notions of \textit{classical DP} before giving the formulation.
        \begin{Definition}[neighboring datasets\textup{~\cite{dwork2014algorithmic}}]            \label{Def:neighbor}
            For two datasets $X,X'\in \mathcal{D}$, we call them neighbors if the following inequality holds:
            \begin{equation}
                d(X,X') \leq 1,
            \end{equation}
            where the function $d(\cdot)$ calculates the Manhattan or Euclidean distance between datasets. 
        \end{Definition}
         Here, neighbors $X$ and $X'$ differ by one entry, which originally describes the subtle change in datasets. The following definition of global sensitivity represents an upper bound of a function (or query) on neighbors.
         \begin{Definition}[global sensitivity\textup{~\cite{dwork2014algorithmic}}]
         \label{Def:sens}
             For two neighboring datasets $X,X'\in \mathcal{D}$, and a function (or query) $f:\mathcal{D}\rightarrow \mathcal{S}$, the global sensitivity $\Delta f$ is the least upper bound of the distance between neighbors with $f$ applied. Taking the Manhattan distance as an example: $\Delta f = \underset{X,X';d(X,X')\leq 1}{\max} \left \| f(X)-f(X') \right \|_1$.
         \end{Definition}
         Definition~\ref{Def:sens} indicates the maximum change in the function (or query)'s output when its input changes slightly (by addition or removal of one entry). In general, a higher sensitivity requires a greater mount of randomization to ensure DP.
         
         The DP property of datasets characterizes the indistinguishability of query outputs under the neighboring setting. Using Def. \ref{Def:neighbor} and Def. \ref{Def:sens}, the formal definition of DP can be obtained, as proposed in a series of published articles~\cite{dwork2006differential,dwork2006calibrating,dwork2010boosting,dwork2014algorithmic}.
        \begin{Definition}[$(\epsilon,\delta)$-DP]
        \label{Def:classical_dp}
            A randomized algorithm $\mathcal{M}$ satisfies $(\epsilon,\delta)$-DP if for any pair of neighboring datasets $X,X'\in \mathcal{D}$ and all possible outcomes $\mathcal{S}$, then we have
        \begin{equation}
		  \mathrm{Pr}[\mathcal{M}(X)\in \mathcal{S}] \leq \exp(\epsilon) \cdot \mathrm{Pr}[\mathcal{M}(X')\in \mathcal{S}]  + \delta,
          \label{eq:dp}
	\end{equation}
        where \equref{eq:dp} holds with probability at least $1-\delta$. We define $\mathcal{M}$ to be $\epsilon$-DP by setting $\delta=0$.
        \end{Definition}
        
        In the classical scenario, a DP mechanism is usually realized by applying a perturbation directly onto the input data, query output~\cite{dwork2006differential,dwork2010boosting} or gradients~\cite{abadi2016deep}, such as the Laplace mechanism and the Gaussian mechanism.
        \begin{Definition}[additive noise mechanism]
            \label{Def:addnoise}
            Given a function (or query) $f:\mathcal{D}\rightarrow \mathcal{S}$ and a noise distribution $\mathcal{N}$, the additive noise mechanism can be described as
            \begin{equation}
                \mathcal{M}_{\mathcal{N}}(X,f(\cdot),\epsilon,\delta) = f(X) + \mathcal{N}(0,\sigma^2).
            \end{equation}
        \end{Definition}
       Definition \ref{Def:addnoise} instantiates the Laplace mechanism when $\mathcal{N}\sim \mathrm{Lap}(0,\Delta f/\epsilon)$, where $\mathrm{Lap}(\mu,\sigma) = \frac{1}{2\sigma}\exp(-\frac{|x-\mu|}{\sigma})$, and the Gaussian mechanism when $\mathcal{N}\sim \mathrm{Gau}(0,\sqrt{2\ln(1.25/\delta)}\cdot\Delta f/\epsilon)$, where $\mathrm{Gau}(\mu, \sigma) = \frac{1}{\sqrt{2\pi}\sigma}\exp(-\frac{(x-\mu)^2}{2\sigma^2})$. $\epsilon$, $\delta$ and $\Delta f$ are the privacy parameters and the global sensitivity introduced in \defref{Def:classical_dp} and \defref{Def:sens}, respectively.
        
        In contrast to directly adding noise to a range of one dataset, the randomized response mechanism (RR)~\cite{erlingsson2014rappor} presents randomization by flipping biased coins, creating privacy about the true query answer response to one data point. 
        \begin{Definition}[randomized response mechanism]
            \label{def:rr}
            Suppose that a function (or query) $f:\mathcal{D}\rightarrow \mathcal{S}$ returns a true answer with a probability of the coin being a head of $\frac{\exp(\epsilon)}{\exp(\epsilon)+1}$, where $\epsilon\in[0, \infty]$. Then, the randomized response mechanism can be described as
        \begin{equation}
            \label{eq:rrDPm}            
            \mathcal{M}_{\mathrm{RR}}(X, f(\cdot),\epsilon) = \frac{\exp(\epsilon)}{\exp(\epsilon)+1} f(X) + \frac{1}{\exp(\epsilon)+1} \tilde{f}(X),
        \end{equation}
        where $\tilde{f}(\cdot)$ returns a random answer.
        \end{Definition}
        The implementation of the RR mechanism implies a stronger version of DP, namely local differential privacy (LDP).
        \begin{Definition}[$(\epsilon,\delta)$-LDP]
            \label{def:ldp}
            A randomized algorithm $\mathcal{M}$ satisfies $(\epsilon,\delta)$-LDP if for any pair of datapoints $x,x'$ sampled from $\mathcal{D}$ and the possible outcome $y$, then we have
        \begin{equation}
		  \mathrm{Pr}[\mathcal{M}(x)=y] \leq \exp(\epsilon) \cdot \mathrm{Pr}[\mathcal{M}(x')=y] + \delta,
          \label{eq:ldp}
        \end{equation}
        where \equref{eq:ldp} holds with probability at least $1-\delta$. We define $\mathcal{M}$ to be $\epsilon$-LDP by setting $\delta=0$.
        \label{Def:ldp}
        \end{Definition}
        Notably, LDP considers the information that a sample is distinguished from another sample as private, rather than a handful of samples, which means that there is no need to impose~\defref{Def:neighbor} constraint on it.

\section{Differential privacy\\ from classical extension to quantum}
    \label{sec:qdp}
\begin{table*}  
                \centering
                \caption{Different distance metrics introduced in QDP publications. Two quantum states are denoted as $\rho$ and $\sigma$.}
                \vspace{-3pt}
                \label{tab:dis_metric}
                \scriptsize
                \begin{tabularx}{\textwidth}{p{2.5cm}| X |p{9.5cm}}
                    \toprule
                    Reference (cites) & Metric & Description  \\
                    \midrule
                    Zhou et al.~\cite{zhou2017differential} & bounded trace distance & Trace distance between two quantum states is represented as $\tau_{\mathrm{tr}}(\rho,\sigma)=\left \| \rho-\sigma \right \|_{\rm{tr}} = \rm{Tr}(|\rho-\sigma|)/2=\mathrm{Tr}(\sqrt{(\rho-\sigma)^\dagger(\rho-\sigma)})/2$, bounded with $d$, that is $\tau_{\mathrm{tr}}(\rho,\sigma)\leq d$. \\   
                    \midrule
                    Zhou et al.~\cite{zhou2017differential} & proportional distance & For all POVMs $\{M_i\}_{i\in\mathcal{O}}$, the proportional distance is described as the maximum log of proportion of the post-measurement probabilities when $\{M_i\}_{i\in\mathcal{O}}$ are performed on $\rho,\sigma$, that is $\tau_{\mathrm{PD}}(\rho,\sigma)=\sup_{\{M_i\}_{i\in\mathcal{O}}}\max\{\ln(\mathrm{Tr}(M_i\rho)/\mathrm{Tr}(M_i\sigma)),\ln(\mathrm{Tr}(M_i\sigma)/\mathrm{Tr}(M_i\rho))\}$. \\  
                    \midrule
                    Gong et al.~\cite{gong2022enhancing} & normalized Hamming distance & $\rho$ and $\sigma$ can be represented as $\rho=e_\rho |0\rangle^{\otimes n}$ and $\sigma=e_\sigma |0\rangle^{\otimes n}$ via encoding from initial state $|0\rangle^{\otimes n}$, then the normalized Hamming distance between states is $\tau_{\mathrm{ham}}(\rho,\sigma)=1/n\sum^n_{i=1}\mathbf{1}[e_\rho^i\neq e_\sigma^i]$, bounded with $d$, that is $\tau_{\mathrm{ham}}(\rho,\sigma)\leq d$.\\
                    \midrule
                    Hirche et al.~\cite{hirche2023quantum} &  Wasserstein distance of order $1$ & Wasserstein distance of order 1 between $\rho$ and $\sigma$ is represented as $\tau_{W_1}(\rho,\sigma)={\sup}_{\{M_i\}_{i\in\mathcal{O}}} \mathrm{Tr}(M_i\rho-M_i\sigma)$, bounded with $d$, that is, $\tau_{W_1}(\rho,\sigma)\leq d$.\\
                    \midrule
                    Aaronson et al.~\cite{aronson2019gentle} & reachability by single-register operation & $\rho$ and $\sigma$ are neighbors if each of them reaches either ($\sigma$ from $\rho$, or $\rho$ from $\sigma$) by performing a super-operator on a single qubit only. \\
                    \midrule
                    Angrisani et al.~\cite{angrisani2023unifying} & $(\mathcal{L},d)$-neighboring & $\rho$ and $\sigma$ are $(\mathcal{L},d)$-neighboring if $\exists \ell\in\mathcal{L}$ and $\mathcal{L} \subset P([n])$, $\mathrm{Tr}(M_\ell \rho)=\mathrm{Tr}(M_\ell\sigma) \wedge \left \|\rho-\sigma\right \|_{\rm{tr}}\leq d$. If $\mathcal{L}=\{[n]\}$, it is equivalent to trace distance bounded with $d$. If $\mathcal{L}=1$ and $d=1$, it recovers the metric that reachability by single register operation.  \\
                    \bottomrule
                \end{tabularx}
            \end{table*}
    In recent years, researchers have attempted to transfer classical algorithms or mechanisms to quantum scenarios~\cite{rebentrost2014quantum,zhou2017differential,cong2019quantum,verdon2019learning,bausch2020recurrent}, one of which is the generalization of DP (\defref{Def:classical_dp}) to the quantum domain~\cite{zhou2017differential}. In this section, we introduce how to extend the classical DP definition into quantum settings.
    
    %Inspired by the line of formulation of classical DP (Section~\ref{subsec:formal_cdp}), the following describes how to extend DP to a quantum version.
    
\subsection{The formulation of quantum DP}
\label{subsec:qdp}
     In QC, information is typically encoded into quantum states, which may include sensitive information. The collapse of states also accidentally leaks privacy if the measurement-extracted information is used for training. Consequently, it is imperative to provide privacy preservation for the process of QC, especially for these quantum states. To adapt classical DP principles to quantum settings, it is necessary to first establish the concept of neighboring relationships between quantum states. Like neighboring datasets in classical DP (\defref{Def:neighbor}), various distance metrics have been introduced to quantify the distances between quantum states. The formal definition of neighboring quantum states is presented as follows.
        \begin{Definition}[neighboring quantum states]
            \label{def:neighbor_states}
            For two quantum states $\rho,\sigma\in \mathcal{H}$, we call them neighbors within distance $d$ if the following inequality holds:
            \begin{equation}
                \tau(\rho,\sigma) \leq d,
            \end{equation}
            where $\tau(\cdot)$ is denoted as a kind of distance metric.
        \end{Definition}

    Different kinds of distance metrics introduced in QDP publications are given in~\tabref{tab:dis_metric}. Zhou et al.~\cite{zhou2017differential} first state their definition of the quantum version of "neighboring". The closeness of two close quantum states is measured by trace distance, that is, $\left \| \rho-\sigma \right \|_{\rm{tr}} = \rm{Tr}(|\rho-\sigma|)/2$. In addition, they propose the proportional distance to measure the relationship between states via the maximum proportion of both post-measurement probabilities. Aaronson and Rothblum introduce a classical-like metric \cite{aronson2019gentle}, where two quantum states are neighbors if any state can reach another state ($\sigma$ from $\rho$ or $\rho$ from $\sigma$) by performing a channel on a single qubit only. If quantum states are described by tensor products, Gong et al.~\cite{gong2022enhancing} present the normalized Hamming distance to measure the closeness between states. Inspired by the work of~\cite{de2021quantum}, Hirche et al.~\cite{hirche2023quantum} and Angrisani et al.~\cite{angrisani2023unifying} propose the quantum Wasserstein distance instead of the trace distance to obtain tighter privacy upper bounds. A generalized quantum neighboring relationship is presented in~\cite{angrisani2023unifying}, with trace distance and reachability by a single register operation as its two particular cases. Trace distance is the most common metric used to measure the distinguishability of two quantum states. 
    
    On the basis of~\defref{def:neighbor_states}, these works~\cite{zhou2017differential,aronson2019gentle,hirche2023quantum,nuradha2023quantum,angrisani2023unifying} formalize the mathematical meaning of \textit{quantum differential privacy (QDP)} and clarify the following definition. 
        \begin{Definition}[$(\epsilon,\delta)$-QDP]
            \label{Def:QDP}
            A noisy quantum algorithm $\mathcal{A}=(\phi, \mathcal{E},\{M_i\}_{i\in \mathcal{O}})$ satisfies $(\epsilon,\delta)$-QDP if for any pair of input neighboring states $\rho, \sigma$ within distance $d$ and all possible measurement outcomes $\mathcal{O}$, we have
            \begin{equation}
            \sum_{i\in \mathcal{O}}\mathrm{Tr}[M_i\mathcal{E}(\rho)] \leq \exp{(\epsilon(d))} \cdot \sum_{i\in \mathcal{O}}\mathrm{Tr}[M_i\mathcal{E}(\sigma)] + \delta,
            \label{eq:qdp}
            \end{equation}
         where~\equref{eq:qdp} holds with probability at least $1-\delta$. We define $\mathcal{A}$ as $\epsilon$-QDP by setting $\delta=0$.
         \end{Definition}
     
        \begin{figure*}
        \centering
        \includegraphics[width=\linewidth]{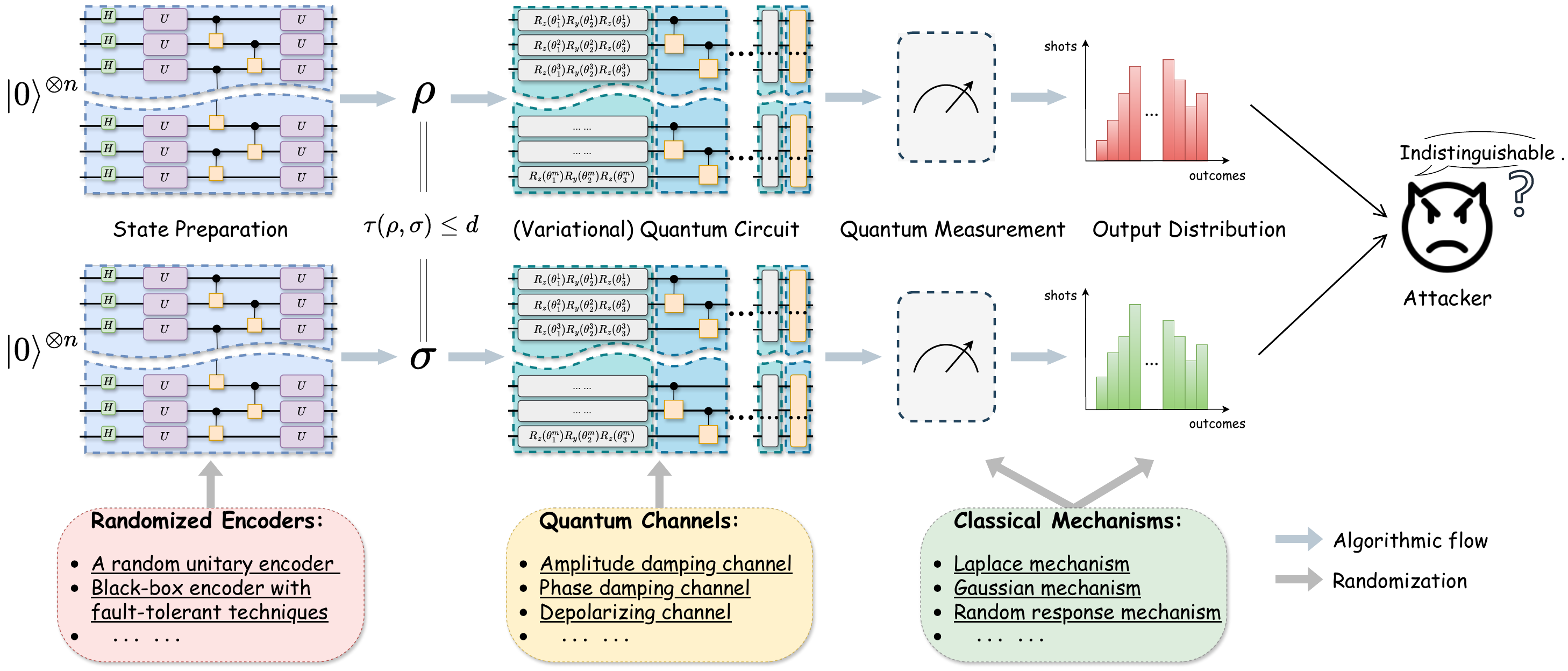}
        \caption{(Q)DP of quantum algorithm.}
        \label{fig:qdp}
        \end{figure*}    
        As depicted in~\figref{fig:qdp}, a quantum algorithm $\mathcal{A}$ is said to possess QDP if, for any pair of quantum states (denoted as $\rho$ and $\sigma$) within distance $d$, the distribution of the algorithm's corresponding measurement output is indistinguishable for the attacker. Referring to related publications, ways to achieve QDP vary greatly, primarily by implementing global or local, internal or external randomization in the quantum algorithm $\mathcal{A}$, which is what Section~\ref{Sec:DP_QA} discusses. 

        Two significant properties of DP are \textit{post-processing} and \textit{Composability}~\cite{dwork2014algorithmic}, which also pertain to QDP. The following theorems describe these properties, respectively.
        \begin{Theorem}[Post-processing\textup{~\cite{dwork2014algorithmic,zhou2017differential}}]
        \label{theorem:postprocess}
            If a quantum algorithm $\mathcal{A}=(\phi, \mathcal{E},\{M_i\}_{i\in \mathcal{O}})$ satisfies $(\epsilon,\delta)$-QDP, then for an arbitrary classical mechanism or quantum operation $\mathcal{F}$, $\mathcal{A}_{\mathcal{F}}=\mathcal{F}\circ \mathcal{A}$ also satisfies $(\epsilon,\delta)$-QDP.
        \end{Theorem}
        \begin{Theorem}[Composition\textup{~\cite{hirche2023quantum}}]
        \label{theorem:composition}
            If quantum algorithms $\mathcal{A}_1$ and $\mathcal{A}_2$ satisfy $(\epsilon_1,\delta_1)$-QDP and $(\epsilon_2,\delta_2)$-QDP, respectively, then $\mathcal{A}_1 \otimes \mathcal{A}_2$ satisfies $(\epsilon,\delta)$-QDP, where $\epsilon=\epsilon_1+\epsilon_2$ and $\delta=\min\{\delta_1+\exp(\epsilon_1)\delta_2,\delta_2+\exp(\epsilon_2)\delta_1\}$.
        \end{Theorem}

    \subsection{Quantum noise}
    \label{subsec:quantum_noise}
        Noise is bound to arise in NISQ devices because of imperfect control in the quantum system and interaction with the environment. Generally, quantum noise can be categorized as coherent or incoherent. The former is unitary and easy to simulate, such as rotation noise, whereas the latter evolves non-unitarily and is what we call the \textit{quantum channel}.
        \begin{comment}
        \begin{table*}
                \centering
                \caption{Categorization of quantum noise models into whether they are coherent or incoherent.}
                \vspace{-3pt}
                \label{tab:quennoise}
                %\scriptsize
                \begin{tabular}{cc}
                    \toprule
                    Whether coherent or not & noise models  \\
                    \midrule
                    coherent & Rotation noise    \\
                    \midrule
                    incoherent &  \makecell{Pauli noise (bit/phase flip/bit-phase and depolarizing) \\ phase/amplitude damping} \\
                    \bottomrule
                \end{tabular}
            \end{table*}
        \end{comment}

         The quantum channel is mathematically modeled as a \textit{completely positive and trace-preserving (CPTP)} linear mapping. Such a mapping $\mathcal{E}$ is known as a superoperator~\cite{nielsen2010quantum}, which is expressed as a set of Kraus matrices $\mathcal{E}=\{E_i\}_{i\in \mathcal{I}}$. The state transformation via the quantum channel can be represented as $\mathcal{E}(\rho)=\sum_i^\mathcal{I}E_i\rho E_i^\dagger$, where $\rho$ is the density matrix form of the input state.

         Typical quantum channels include \textit{Pauli channel} and \textit{damping channel}. Among them, according to the flip probabilities of different Pauli gates, Pauli channels can be divided into bit-flip, phase-flip, bit-phase-flip, and depolarizing channels; according to the description perspective of whether the loss of information or energy in a quantum system, damping channels can be divided into phase-damping channels and amplitude-damping channels.

        \begin{Theorem}[contractility\textup{~\cite{nielsen2010quantum}}]
        \label{theorem:contractility}
            For any pair of quantum states $\rho$ and $\sigma$, and a quantum channel $\mathcal{E}$, here we have
            \begin{equation}
                \tau(\mathcal{E}(\rho),\mathcal{E}(\sigma)) \leq \tau(\rho,\sigma).
            \end{equation}
        \end{Theorem}
        \theref{theorem:contractility} indicates that the distance between states decreases after being transformed by the quantum channel.

        The unavoidable inherent noise in NISQ devices poses computing difficulties but also naturally makes the quantum algorithm differentially private~\cite{du2021quantum}. We next present a case study of how quantum noise can be utilized to obtain QDP.

    \begin{table*}
        \centering 
        \caption{DP-preserving quantum algorithm in selective publications that are mentioned in Section~\ref{sec:DP_qa}. Assume that there is a raw quantum algorithm $\mathcal{A}=(\phi,\mathcal{E},M)$. $\mathcal{N}$, $\mathrm{RR}$ and $\mathcal{R}$ are randomization mechanisms, instantiated by~\defref{Def:addnoise},~\defref{def:rr} and~\defref{Def:quantum_pri_mechansim}. According to Section~\ref{subsec:quantum_noise}, quantum noise can be modeled as Pauli channels or damping channels.}
        
        \label{tab:dp_in_qa}
        \begin{tabular}{c|c|c|c|c}
            \toprule
                Phase    & DP-preserving quantum algorithm $\mathcal{A}$ & Reference (cites) & Type & Implementation  \\ \midrule
                \multirow{3}{*}{\makecell{State preparation}} 
                    & $(\phi,\mathcal{E},M)$ with $\mathcal{M}_\mathcal{N}(x)$ as input& \cite{senekane2017privacy,yoshida2020classical} & $\epsilon$-DP & Laplace noise\\ 
                    &$(\phi_{\mathcal{N}},\mathcal{E},M)$&\cite{du2022quantum} & $(\epsilon,\delta)$-DP & Gaussian/Laplace noise\\
                    & $(\phi_{\mathcal{R}} \circ \phi,\mathcal{E},M)$&\cite{gong2022enhancing}  & $(\epsilon,\delta)$-QDP & Randomized encoding  \\
                    & $(\phi,\mathcal{E},M)$ if $\phi$ is basic/amplitude encoding&\cite{angrisani2022differential}&$(0,\delta)$-DP & Gaussian noise\\ \midrule
                    \multirow{2}{*}{\makecell{Quantum circuit}} 
                    & $(\phi,\mathcal{E}_{\mathcal{N}}\circ\mathcal{E},M)$& \cite{zhou2017differential,hirche2023quantum,angrisani2023unifying,angrisani2022differential,du2021quantum,guan2023detecting} &$(\epsilon,\delta)$-QDP & Quantum noise\\ 
                    &  $(\phi,\mathcal{E},M)$ with noise on gradients of PQC& \cite{li2021quantum,watkins2023quantum, rofougaran2023federated} & $(\epsilon,\delta)$-DP & Gaussian noise\\ \midrule
                    \multirow{3}{*}{\makecell{Quantum measurement}} & $(\phi,\mathcal{E},M)$ with finite shots of measurements&\cite{li2023differential}&$\epsilon$-QDP & Shot noise\\
                    &  $(\phi,\mathcal{E},\mathcal{M}_{\mathrm{RR}}(M)))$ &\cite{angrisani2022quantum}&$\epsilon$-LDP & Randomized response\\ 
                    & $(\phi,\mathcal{E},\mathcal{M}_{\mathcal{N}}(M))$ & \cite{angrisani2023unifying,aronson2019gentle,angrisani2022quantum}& $(\epsilon,\delta)$-QDP & Gaussian/Laplace noise\\  \bottomrule
        \end{tabular}
    \end{table*}

    \subsection{A case study: How to obtain QDP using quantum noise?}      
    \label{subsec:case_study}
    We first present the definition of the quantum privacy mechanism~\cite{zhou2017differential}, which can be described by a composition of the noisy quantum channel and the noiseless quantum circuit.
    \begin{Definition}[quantum privacy mechanism]
    \label{Def:quantum_pri_mechansim}
        Suppose that there is a quantum circuit $\mathcal{E}$ and a noisy quantum channel $\mathcal{E}_\mathcal{N}$, for an input state $\rho$, the quantum privacy mechanism can be described as
        \begin{equation}
          \label{eq:qn}
        \mathcal{E}_\mathcal{N}\circ \mathcal{E}(\rho):\rho \mapsto \mathcal{E}_\mathcal{N}(\mathcal{E}(\rho)). 
        \end{equation}
    \end{Definition}
        
        The noiseless quantum circuit $\mathcal{E}$ can be directly considered as a black box with an input $\rho$ and an output $\tilde{\rho}=\mathcal{E}(\rho)$. 

        Taking the depolarizing channel as an example, here, $\mathcal{E}_\mathcal{N}=\left\{
            E_0=\sqrt{1-p}I, \ E_1=\sqrt{\frac{p}{3}}X, \ E_2=\sqrt{\frac{p}{3}}Y, \ E_3=\sqrt{\frac{p}{3}}Z\right\}$, where $p$ is the error probability of depolarization. The quantum privacy mechanism that post-processing $\mathcal{E}$ with $\mathcal{E}_\mathcal{N}$ is 
        \begin{equation}
            \mathcal{E}_\mathcal{N}(\tilde{\rho}) = \sum_{i=0}^3E_i\tilde{\rho}E_i^\dagger=\frac{pI}{D} + (1-p)\tilde{\rho},
        \end{equation}
        where $D$ is the dimension of the quantum system~\cite{nielsen2010quantum}.

        For neighboring quantum states $\rho$ and $\sigma$ within distance $d$, the outputs of the circuit are $\tilde{\rho}=\mathcal{E}(\rho)$ and $\tilde{\sigma}=\mathcal{E}(\sigma)$. We subsequently examine the neighboring relationship between output states with $\tau(\tilde{\rho},\tilde{\sigma})\leq d$ (\textit{cf.}~\theref{theorem:contractility}).

        We can bootstrap the notion of \textit{privacy loss} from~\defref{Def:QDP} by analogy classical cases~\cite{dwork2014algorithmic}. For a quantum algorithm $\mathcal{A}=(\phi, \mathcal{E},\{M_i\}_{i\in \mathcal{O}})$ and neighboring quantum states $\rho$ and $\sigma$, the privacy loss is defined as
            \begin{equation}
                \label{eq:pri_loss}
                \mathcal{L}_{\mathcal{A}(\rho)\parallel \mathcal{A}(\sigma)} =\ln \frac{\sum_{i\in \mathcal{O}}\mathrm{Tr}[M_i\mathcal{E}(\rho)]}{\sum_{i\in \mathcal{O}}\mathrm{Tr}[M_i\mathcal{E}(\sigma)]}.
            \end{equation}

        According to~\equref{eq:qdp} and~\equref{eq:pri_loss}, for a set of arbitrary measurements $M$, we observe
        \begin{equation}
        \label{eq:casestudy}
            \begin{aligned}
            \epsilon &\leq \ln \frac{\sum_{i\in \mathcal{O}}\mathrm{Tr}[M_i\mathcal{E}_\mathcal{N}(\tilde{\rho})]}{\sum_{i\in \mathcal{O}}\mathrm{Tr}[M_i\mathcal{E}_\mathcal{N}(\tilde{\sigma})]} 
            = \ln\frac{\sum_{i\in \mathcal{O}}\mathrm{Tr}[M_i\sum_{i=0}^3E_i\tilde{\rho}E_i^\dagger]}{\sum_{i\in \mathcal{O}}\mathrm{Tr}[M_i\sum_{i=0}^3E_i\tilde{\sigma}E_i^\dagger]} \\
            &=\ln\frac{\sum_{i\in \mathcal{O}}\mathrm{Tr}\left[M_i\left(\frac{pI}{D}+(1-p)\tilde{\rho}\right)\right]}{\sum_{i\in \mathcal{O}}\mathrm{Tr}\left[M_i\left(\frac{pI}{D}+(1-p)\tilde{\sigma}\right)\right]} 
            \\&=\ln\left[\frac{(1-p)\sum_{i\in \mathcal{O}}\mathrm{Tr}\left[M_i\left(\tilde{\rho}-\tilde{\sigma}\right)\right]}{\sum_{i\in \mathcal{O}}\mathrm{Tr}\left[M_i\left(\frac{pI}{D}+(1-p)\tilde{\sigma}\right)\right]} + 1\right] 
            \\&\leq \ln\left[\frac{(1-p)d\cdot \sum_{i\in \mathcal{O}}\mathrm{Tr}[M_i]}{\frac{p}{D}\cdot \sum_{i\in \mathcal{O}}\mathrm{Tr}[M_i]} + 1\right] 
            = \ln\left[\frac{1-p}{p}dD+1\right].
            \end{aligned}
        \end{equation}
         This equation is a result of~\cite[Theorem 3]{zhou2017differential}. In conclusion, for a quantum algorithm $\mathcal{A}=(\phi, \mathcal{E}_{\mathcal{N}}\circ\mathcal{E},M)$, we denote that it satisfies $\epsilon$-QDP, where $\epsilon\leq\ln\left[\frac{1-p}{p}dD+1\right]$.

\section{Differential privacy in noisy quantum algorithms}
\label{Sec:DP_QA}

\label{sec:DP_qa}
    As illustrated in Section~\ref{sec:qdp}, the quantum algorithm is orderly composed of the phases of state preparation (encoding and initial states), quantum circuits, and quantum measurements. This section focuses on how internal or external randomization existing in these phases can be utilized to create or enhance privacy. We survey some recent developments, concentrating on the implementations of DP and QDP mechanisms in quantum algorithms. We evaluate their performance in terms of privacy parameters. These findings are summarized in~\tabref{tab:dp_in_qa}. Note that QDP can be realized through DP mechanisms; it is distinguished from DP primarily by its definitions, yet both provide rigorous privacy protection.

    \subsection{Differential privacy preservation in state preparation}
        \label{subsec:sp_dp}
         This first part is dedicated to how DP and QDP are implemented in the phase of state preparation of the quantum algorithm. Assuming that we have a dataset $X$ and an encoding method $\phi$, we have initiated quantum states via~\equref{eq:state_prep}.
         
         Intuitively, we can add noise to the input classical data to guarantee privacy. Senekane et al.~\cite{senekane2017privacy} first discussed implementing the additive noise mechanism (\defref{Def:addnoise}) on classical data, and then transforming the perturbed input into quantum states for utilization in a QML model of binary classification. Let $\tilde{X}$ be randomized by a discrete Laplace distribution $\mathcal{N}$, that is, $\tilde{X} \leftarrow \mathcal{M}_{\mathcal{N}}(X,\epsilon)$. For a quantum algorithm $\mathcal{A}=(\phi,\mathcal{E},M)$ with input $\tilde{x}$, since $\tilde{x}$ is $\epsilon$-DP, $\mathcal{A}$ satisfies $\epsilon$-DP (\textit{cf.} \theref{theorem:postprocess}), but it has very little relevance to QDP, as defined in \equref{eq:qdp}. The same case is called classical-quantum DP in~\cite{yoshida2020classical}, where Yoshida and Hayashi demonstrated that a quantum algorithm with perturbed input costs more on performance and has less advantage than does classical DP.

         On the other hand, Angrisani et al.~\cite{angrisani2022differential} reported that quantum encoding can naturally lead to classical $(\epsilon,\delta)$-DP. To prove this, the notion of a \textit{quantum minimum adjacent kernel} is defined based on~\equref{eq:kernel}, as follows. 
         \begin{Definition}[quantum minimum adjacent kernel\textup{~\cite{angrisani2022differential}}]
         \label{def:qmak}
         For any two datapoints $x,x'$ that are sampled from $\mathcal{D}$, a quantum minimum adjacent kernel is defined as the minimization of the quantum kernel, namely 
         \begin{equation}
         \label{eq:qmak}
             \hat{\kappa} := \underset{d(x,x')\leq 1}{\min} \kappa(x,x').
         \end{equation}
         \end{Definition}
        Let us consider a pair of neighboring datapoints $x$ and $x'$\footnote{In this scenario, $x$ and $x'$ are said to be neighbors if $\rho(x)$ and $\rho(x')$ are neighboring states.}. Their respective encodings are $\rho(x)$ and $\rho(x')$. Reviewing the definition of QDP (\defref{Def:QDP}), if $\epsilon=0$, then the following inequality holds:
         \begin{equation}
         \label{eq:encoding_dp}
             \sum_{i\in \mathcal{O}}\mathrm{Tr}[M_i\mathcal{E}(\rho(x))] - \sum_{i\in \mathcal{O}}\mathrm{Tr}[M_i\mathcal{E}(\rho(x'))] \leq \delta.
         \end{equation}
         According to~\equref{eq:kernel},~\equref{eq:qmak} and~\theref{theorem:contractility}, the left half of the inequality above, denoted as $\Delta$, can be properly scaled, 
         \begin{equation}
             \begin{aligned}
              \Delta &\leq \tau_{\mathrm{tr}}(\mathcal{E}(\rho(x)), \mathcal{E}(\rho(x'))) 
                  \leq \tau_{\mathrm{tr}}(\rho(x), \rho(x')) 
                 \\&= \sqrt{1-|\langle \phi(x) | \phi(x') \rangle|^2} = \sqrt{1-\kappa(x,x')}
                  \leq \sqrt{1-\hat{\kappa}}.
             \end{aligned}
         \end{equation}
         Here, we have $\delta \geq \sqrt{1-\hat{\kappa}}$. Thus the encoding satisfies $(0,\sqrt{1-\hat{\kappa}})$-DP.
         
        A quantum algorithm with this encoding $\phi$ also satisfies classical $(0,\sqrt{1-\hat{\kappa}})$-DP (\textit{cf.} \theref{theorem:postprocess}). Specifically, if $\phi$ is basic encoding, $\hat{\kappa}$ equals $1-\frac{1}{l}$, where $l$ is the length of the dataset; if $\phi$ is amplitude encoding, $\hat{\kappa}$ equals $1-\underset{i}{\max} |x_i|^2$. In addition, there is nearly no privacy guarantee when $\phi$ is rotation encoding because $\hat{\kappa}=0$ in this case, which means that~\equref{eq:dp} holds with probability 0. 

        \begin{table*}[t]
        \centering
        \caption{QDP in noisy quantum circuit. $D$ is the demension of quantum system and $n,k$ are the number of qubits ($D=2^n$). $p$, $\gamma$ and $\lambda$ are the error probability of depolarization, amplitude damping and phase damping, respectively.}
        \label{tab:qdp_qc}
        \begin{tabularx}{\textwidth}{p{4.5cm}| X |p{6.5cm}}
        \toprule
            Reference (cites) & Noise model &  Upper bound of privacy parameters \\
        \midrule    
             Zhou et al.~\cite[Theorem 3]{zhou2017differential} & depolarizing channel  & $\epsilon \leq \ln\left(1+\frac{D}{p}(1-p)d\right)$ \\
          
             Hirche et al.~\cite[Corollary \uppercase\expandafter{\romannumeral4}.3]{hirche2023quantum} & global depolarizing channel &  $\epsilon \leq \max\left\{0,\ln\left(1+\frac{D}{p}((1-p)d-\delta)\right) \right\}$ \\
            
             Hirche et al.~\cite[Corollary \uppercase\expandafter{\romannumeral4}.6]{hirche2023quantum} & local depolarizing channel &  $\epsilon \leq \max\left\{0,\ln\left(1+\frac{2^k}{p^k}((1-p^k)d-\delta)\right) \right\}$ \\
          
             Angrisani et al.~\cite[Corollary 5.1]{angrisani2023unifying} & global generalized noisy channel & $\delta\leq \max\left\{ 0, (1-\exp(\epsilon))\frac{p}{D}+(1-p)d\right\}$\\
           
             Angrisani et al.~\cite[Corollary 5.2]{angrisani2023unifying} & local generalized noisy channel & $\delta\leq \max\left\{ 0, (1-\exp(\epsilon))\frac{p^k}{2^k}+(1-p^k)d\right\}$ \\
             %phase damping  &$\mathcal{E}_{pd}$& \\
            
             Zhou et al.~\cite[Theorem 1]{zhou2017differential} & generalized amplitude damping channel  & $\epsilon \leq \ln\left(1+\frac{2d\sqrt{1-\gamma}}{1-\sqrt{1-\gamma}}\right)$ \\
         
             Zhou et al.~\cite[Theorem 2]{zhou2017differential} & phase-amplitude damping channel & $\epsilon \leq \ln\left(1+\frac{2d\sqrt{1-\gamma}\sqrt{1-\lambda}}{1-\sqrt{1-\gamma}\sqrt{1-\lambda}}\right)$\\
         
             Angrisani et al.~\cite[Theorem 7]{angrisani2022differential} & depolarizing-phase-amplitude damping channel &  $\epsilon \leq (1-p)\ln\left(1+\frac{2d\sqrt{1-\gamma}\sqrt{1-\lambda}}{1-\sqrt{1-\gamma}\sqrt{1-\lambda}}\right)$\\
             %Hirche et al.~\cite[ \uppercase\expandafter{\romannumeral4}.21]{hirche2023quantum} & arbitrary local qubit noisy channel &  $\delta = \left(\frac{1}{2}\sqrt{(1+\exp(\epsilon))^2-4\exp(\epsilon)\left(\frac{\lambda_{\text{min}}(C_{\mathcal{N}^\dagger\circ \mathcal{N}})}{4}\right)^k} + \frac{1-\exp(\epsilon)}{2} \right)^n d$ \\ 
        \bottomrule
        \end{tabularx}
    \end{table*}

        Since quantum encoding naturally leads to classical DP, composing quantum encoding with additive noise intuitively may induce privacy amplification. In~\cite[Algorithm 1]{angrisani2022differential}, the quantum states are measured and then randomized directly via the additive noise mechanism. $\Delta f$ in \defref{Def:addnoise} is set equal to $\sqrt{1-\hat{\kappa}}+t$ for any $t\leq 0$, and the remaining parameters (including $\epsilon$ and $\delta$) are tuned independently. Using the Laplace and Gaussian distributions, $(0,\sqrt{1-\hat{\kappa}})$-DP is scaled up to $\epsilon$-DP and $(\epsilon,\delta)$-DP, with probability at least $1-4\exp(-|\mathcal{O}|\cdot t^2)$, where $|\mathcal{O}|$ is the size of the set of all possible outcomes. The full proof can be found in~\cite[Theorem 2 \& 3]{angrisani2022differential}.

        However, both methods of manually adding noise to the input classical dataset or selecting a specific quantum encoding method only satisfy classical DP (\defref{Def:classical_dp}) but not QDP (\defref{Def:QDP}). Some researchers~\cite{du2022quantum, gong2022enhancing} considered randomized encodings as a perturbation for quantum states, thereby realizing QDP in the phase of state preparation. Du et al.~\cite{du2022quantum} perturbed the initial states directly via Gaussian or Laplace noise. Gong et al.~\cite{gong2022enhancing} applied extra randomized encodings that would generate barren plateaus of gradients when updating parameters, thus improving adversarial robustness and leading to QDP. Note that a codebook is introduced in~\cite{gong2022enhancing}, with various randomized encodings, including a white-box and a black-box. Under the assumption that a quantum algorithm satisfies $\epsilon$-QDP, white-box encodings (e.g., random unitary encoding) that satisfy the 2-design property~\cite{dankert2009exact} can induce $(\epsilon(d),\delta)$-QDP, while the privacy strength varies linearly with $d$. Black-box encodings (e.g., quantum error correction (QEC) encoding) that encode each logical qubit into $n_0$ physical qubits lead to a privacy amplification and satisfy $\left(\epsilon\left(\frac{n_0(n_0-1)d^2}{\delta} \right),\delta\right)$-QDP.

    \subsection{Differential privacy preservation in a quantum circuit}
        \label{subsec:qc_dp}
        This part is dedicated to the implementations of the DP and QDP mechanisms in the quantum circuit. As introduced in Section~\ref{subsec:quantum_noise}, incoherent noise in circuits is unavoidable in quantum algorithms running on NISQ devices. Zhou and Ying~\cite{zhou2017differential} first introduced the concept of QDP by directly considering different quantum channels as quantifiable noise. They built three different noise models based on channels, i.e., depolarizing, amplitude damping, and phase damping, each with a corresponding real-world physical realization. Zhou and Ying generalized the notion of classical DP to the quantum domain by post-processing a black-box quantum circuit with a noisy quantum channel, as introduced in~\defref{Def:quantum_pri_mechansim}. They derived upper bounds of privacy loss for the generalized amplitude damping mechanism, phase-amplitude damping mechanism, and depolarizing mechanism, respectively (see more details in \tabref{tab:qdp_qc}). Referring to~\cite{dwork2014algorithmic}, they proved that their concept of QDP satisfies a series of composition theorems (\theref{theorem:composition}), including advanced composition.

        Hirche et al.~\cite{hirche2023quantum} explained the concept of QDP based on information-theoretic approaches and obtained a much tighter bound than did the above methods~\cite{zhou2017differential}, as shown in~\tabref{tab:qdp_qc}. Specifically, they built several noise models of QDP, such as global and local depolarizing channels, based on the \textit{quantum hockey-stick divergence} and \textit{smooth max-relative entropy}. 
        \begin{Definition}[quantum hockey-stick divergence\textup{~\cite{hirche2023quantum}}]
        \label{Def:hockey}
            For any two quantum states $\rho, \sigma$, the hockey-stick divergence of $\rho$ with respect to $\sigma$ is defined as
            \begin{equation}
                D_{\alpha}(\rho \parallel \sigma):=\mathrm{Tr}(\rho-\alpha\sigma)^+, \quad  \forall  \alpha \geq 1.
            \end{equation}
        \end{Definition}
        Note that $D_{1}(\rho \parallel \sigma)$ is equivalent to its trace distance.
         
         Like \theref{theorem:contractility}, quantum hockey-stick divergence is also contracted; thus, the notion of the \textit{contraction coefficient} of quantum hockey-stick divergence is defined as 
        \begin{equation}
        \label{eq:contraction_coefficient}
            \eta_\alpha(\mathcal{E})=\underset{\rho,\sigma}{\sup} \frac{D_\alpha(\mathcal{E}(\rho) \parallel \mathcal{E}(\sigma))}{D_\alpha(\rho\parallel\sigma)},
        \end{equation}
        where $\mathcal{E}$ is a quantum channel. 
        
        The $\delta$-ball of subnormalized quantum states around $\mathcal{H}$ is defined via the trace distance as $B^\delta(\rho)=\{\hat{\rho}|\hat{\rho}\in\mathcal{H}\wedge \tau_{\mathrm{tr}}(\rho,\hat{\rho}) \leq \delta\}$.
        Here, we have the following definition.
         \begin{Definition}[smooth max-relative entropy\textup{~\cite{hirche2023quantum}}]
         \label{Def:mre}
              For any two quantum states $\rho, \sigma$, the smooth max-relative entropy of $\rho$ with respect to $\sigma$ is defined as
              \begin{equation}
                  E_{\max}^\delta(\rho \parallel \sigma):=\underset{\hat{\rho}\in B^\delta(\rho)}{\inf}E_{\max}(\hat{\rho}\parallel \sigma),
              \end{equation}
              where $E_{\max}(\hat{\rho}\parallel \sigma)=\inf\{\lambda\in\mathbb{R}:\hat{\rho}\leq2^\lambda\sigma\}$ is the max-relative entropy.
         \end{Definition}
         \cite[Lemma III.2]{hirche2023quantum} presented the sufficient and necessary condition between~\defref{Def:hockey} and~\defref{Def:mre}. For any pair of neighboring quantum states $\rho,\sigma$, we have 
         \begin{equation}
         \label{eq:snc}
             E_{\max}^\delta(\rho \parallel \sigma) \leq \ln \alpha \Leftrightarrow D_\alpha(\rho \parallel \sigma)\leq \delta.
         \end{equation}
        This equivalence condition connects $\delta$ with privacy loss (if set $\alpha=\exp(\epsilon)$) so that we can obtain a tighter upper bound of the privacy parameters of QDP. In contrast, the upper bound of a quantum circuit with the depolarizing channel from Zhou's work~\cite{zhou2017differential} is based on the original definition of privacy loss (\equref{eq:pri_loss}), as~\equref{eq:casestudy} does. To formalize \equref{eq:qdp}, $\alpha$ is set equal to $\exp(\epsilon)$. Let $\mathcal{A}=(\phi,\mathcal{E},M)$ be a quantum algorithm. For neighboring states $\rho,\sigma$ with $\tau_{\mathrm{tr}}(\rho,\sigma)\leq d$ and a noise model $\mathcal{E}_{\mathcal{N}}$, $\mathcal{A}_{\mathcal{E}_{\mathcal{N}}}=(\phi,\mathcal{E}_{\mathcal{N}}\circ\mathcal{E},M)$ is said to be $(\epsilon,\delta)$-QDP. According to~\equref{eq:contraction_coefficient} and~\cite[Lemma II.4]{hirche2023quantum}, we have
        \begin{equation}
            \begin{aligned}
                D_{\exp(\epsilon)}(\mathcal{E}_{\mathcal{N}}(\rho) \parallel \mathcal{E}_{\mathcal{N}}(\sigma)) &\leq \eta_{\exp(\epsilon)}(\mathcal{E}_{\mathcal{N}}) \cdot D_{\exp(\epsilon)}(\rho \parallel \sigma) 
                \\
                \leq \eta_{\exp(\epsilon)}(\mathcal{E}_{\mathcal{N}})\cdot D_1(\rho \parallel \sigma) &\leq \eta_{\exp(\epsilon)}(\mathcal{E}_{\mathcal{N}}) \cdot d.
            \end{aligned}
        \end{equation}
        Hence, the privacy parameter $\delta=\eta_{\exp{(\epsilon)}}(\mathcal{E}_{\mathcal{N}})\cdot d$. When the noise model $\mathcal{E}_{\mathcal{N}}$ is a global or local depolarizing channel, the bound of $\epsilon$ is depicted in~\tabref{tab:qdp_qc}.

        Angrisani et al.\cite{angrisani2023unifying} provided more generalized quantum channels for QDP than did~\cite{hirche2023quantum}, and improved the privacy bound via the \textit{joint convexity} of the quantum hockey-stick divergence, which is shown as follows.
        \begin{Theorem}[joint convexity\textup{~\cite{angrisani2023unifying}}]
        \label{theorem:ajc}
        For a triple of quantum states $\rho,\sigma,\varphi$ and $\alpha' = 1 + (1-p)(\alpha-1)$, we have
        \begin{equation}
        \begin{aligned}
            &D_{\alpha'}\left((p\rho+(1-p)\sigma)\parallel (p\rho+(1-p)\varphi)\right) 
            \\ &\leq (1-p)(1-\frac{\alpha'}{\alpha})D_\alpha(\rho\parallel\sigma)+(1-p)\frac{\alpha'}{\alpha}D_\alpha(\sigma\parallel\varphi). 
        \end{aligned}
        \end{equation}
        \end{Theorem}
        Let $\mathcal{E}$ be an arbitrary quantum channel; the corresponding noisy channel can be denoted as $\mathcal{E}_{\mathcal{N}}=p\frac{\mathbb{I}}{2^n}+(1-p)\mathcal{E}$, where $n$ is the number of qubits and where $p\in[0,1]$. Specifically, if $n=1$, $\mathcal{E}_{\mathcal{N}}$ is a Pauli channel; if $\mathcal{E}=\mathbb{I}$, $\mathcal{E}_{\mathcal{N}}$ degrades to a depolarizing channel. For a pair of states $\rho,\sigma$ with trace distance $\tau_{\mathrm{tr}}(\rho,\sigma)\leq d$, if the noisy channel $\mathcal{E}_{\mathcal{N}}$ is applied to the states, then by~\theref{theorem:ajc}, it can be concluded that
         \begin{equation}
         \label{eq:dalpha1}
          \begin{aligned}
             &D_{\alpha'}(\mathcal{E}_{\mathcal{N}}(\rho)\parallel\mathcal{E}_{\mathcal{N}}(\sigma)) \leq \delta
             \\ &\leq(1-p)(1-\frac{\alpha'}{\alpha})D_\alpha(\rho\parallel\frac{\mathbb{I}}{2^n})+(1-p)\frac{\alpha'}{\alpha}D_\alpha(\rho\parallel\sigma),
            \end{aligned}
         \end{equation}
         where $\alpha\geq 1$ and $\alpha' = 1 + (1-p)(\alpha-1)$.

         Similarly, if $\mathcal{E}_{\mathcal{N}}$ is a depolarizing channel as a special case, and $\forall \alpha\geq 1$,
         \begin{equation}
         \label{eq:dalpha2}
         \begin{aligned}
             &D_{\alpha'}(\mathcal{E}_{\mathcal{N}}(\rho)\parallel\mathcal{E}_{\mathcal{N}}(\sigma)) \leq \delta
             \\ &\leq \max\left\{0,(1-\alpha)\frac{p}{2^n}+(1-p)D_\alpha(\rho\parallel\sigma))\right\}.
             \end{aligned}
         \end{equation}
         Combining~\equref{eq:dalpha1} with~\equref{eq:dalpha2} and setting $\alpha=\exp(\epsilon)$, a tighter upper bound of $\delta$ is derived, as shown in~\tabref{tab:qdp_qc}. 

        Both~\cite{hirche2023quantum} and~\cite{angrisani2023unifying} investigated the contraction of the quantum hockey-stick divergence; \theref{theorem:contractility} also indicated that applying a quantum channel on states would contract distance. Let $\mathcal{T}$ be a quantum channel, $\mathcal{T}$ is called \textit{$\beta$-Dobrushin}~\cite{angrisani2022differential} for any pair of states $\rho,\sigma$: 
        \begin{equation}
            \underset{\rho,\sigma}{\sup}\frac{\tau_{\mathrm{tr}}(\mathcal{T}(\rho),\mathcal{T}(\sigma))}{\tau_{\mathrm{tr}}(\rho,\sigma)} \leq \beta.
        \end{equation}
        The definition of QDP (\defref{Def:QDP}) demonstrates that the privacy budget and the upper bound $d$ of distance $\tau_{}(\rho,\sigma)$ are positively correlated. When a QDP-satisfying circuit $\mathcal{E}$ is post-processing with $\mathcal{T}$, the composition $\mathcal{T}\circ\mathcal{E}$ also satisfies QDP (\textit{cf.}~\theref{theorem:postprocess}) with a smaller privacy budget. Taking the depolarizing channel being $\mathcal{T}$ as an example, let $\mathcal{E}$ be an amplitude damping channel; the privacy budget is decayed with a factor of $1-p$, where $p$ is the error probability of depolarization, as shown in~\tabref{tab:qdp_qc} (the last line).

        In~\cite{du2021quantum}, the noisy circuit of a quantum algorithm could be modeled by layers composed of parameterized channels $\mathcal{U}$ (e.g., unitary blocks or layers of a quantum neural network) and noise channels $\mathcal{N}$. Therefore, an $n$-layer circuit can be denoted as $\mathcal{E}=\bigoplus_i^n \mathcal{U}_i\circ \mathcal{N}_i$. Because the global depolarizing channel well describes the noise in deep quantum circuits~\cite{vovrosh2021simple}, we could use the layered model to simplify the computation of the noisy circuit (more than one layer), where quantum noise is simulated by incorporating customizable global depolarizing noise into the noiseless circuit.
        
        Assuming that $\mathcal{N}_i$ is an added depolarizing channel, since noise expectation is independent of position in the circuit~\cite{du2021quantum}, all fine-grained noise models $(p_1,p_2,\dots,p_n)$ distributed across quantum circuits can ultimately be normalized to a global noise model with $p^*=1-\prod_{i}^n(1-p_i)$, where $p_i\in[0,1]$ is the error probability of the corresponding channel $\mathcal{N}_i$. Thus, the layered model helps to easily extend the results in~\tabref{tab:qdp_qc}. Taking the depolarizing channel as an example, for quantum circuits modeled by layering, the privacy budget $\epsilon \leq \ln\left(1+\frac{D}{p^*}(1-p^*)d\right)$, where $D$ is the dimension of the quantum system.

        Although the previously described methods achieve QDP by attaching an additional quantum channel behind the circuit layer, Watkins et al.~\cite{watkins2023quantum} directly applied Gaussian noise on quantum gradients in a training PQC. Let $\rho$ be the initiated state, $\mathcal{E}$ be the quantum circuit, and $M$ be the measurements. The objective function is $f(\theta) = \langle\rho |\mathcal{E}^\dagger(\theta) M \mathcal{E}(\theta) | \rho\rangle$. In general, the gradient of the loss function with respect to the parameters of the rotation gate is obtained via the \textit{parameter-shift rule}: $\nabla f(\theta) = \frac{1}{2}\left[f(\theta+\frac{\pi}{2}) - f(\theta-\frac{\pi}{2})\right]$. Similar to~\cite[Algorithm 1]{abadi2016deep}, the gradient calculated from the parameter-shift rule is clipped first, and then applied by an additive noise mechanism with Gaussian noise:
        \begin{equation}
            \nabla f(\theta)^{\mathrm{DP}} = \nabla f(\theta) / \max \{1, \nabla f(\theta)/C\} + \mathcal{N},
        \end{equation}
        where $\mathcal{N}\sim \mathrm{Gau}(0,\sqrt{2\ln(1.25/\delta)}\cdot C/\epsilon\cdot \mathbb{I})$ and where $C$ is a hyperparameter, depending on the range of gradients. Since the privacy assurance comes from the noise on the gradient, this quantum algorithm satisfies classical $(\epsilon,\delta)$-DP, where the privacy parameters are determined by the noise source $\mathcal{N}$. Similarly, Rofougaran et al.~\cite{rofougaran2023federated} and Li et al.~\cite{li2021quantum} extended this privacy method to DP-preserving quantum federated learning.

        For trained PQCs, Huang et al.~\cite{huang2023certified} attempted to add coherent noise by applying rotation gates $R_x$ with randomized angles on each qubit while the parameters are fixed and the QDP bound is theoretically deduced. Similar to~\equref{eq:casestudy}, they presented the upper bound of the privacy budget based on~\equref{eq:pri_loss} and verified their results by running binary classification experiments on NISQ devices, demonstrating that the impact of rotation noise (a single rotation gate on each wire) on quantum algorithm performance is negligible and that this approach is robust.

        Although the majority of research on QDP~\cite{zhou2017differential,hirche2023quantum,angrisani2023unifying,angrisani2022differential,du2021quantum,huang2023certified} emphasizes the upper bound of the privacy parameters of QDP, a recent work~\cite{guan2023detecting} has shifted the focus toward exploring the lower bound of these related parameters. For a noisy quantum circuit $\mathcal{E}(\rho)=\sum_i^\mathcal{D}E_i\rho E_i^\dagger$, if one represents its dual form as $\mathcal{E}^\dagger(M)=\sum_i^\mathcal{D}E_i^\dagger M E_i$, it is apparent that
        \begin{equation*}
            \begin{aligned}
                 \mathrm{Tr}(\mathcal{E}^\dagger(M)\rho) 
                 %&= \mathrm{Tr}(\sum_i^\mathcal{D}E_i^\dagger M E_i\rho) \\&\leq \mathrm{Tr}(\sum_i^\mathcal{D}M E_i\rho E_i^\dagger) =
                 \leq \mathrm{Tr}(M\mathcal{E}(\rho)).
            \end{aligned}
        \end{equation*}
        According to \equref{eq:qdp}, it can be inferred that
        \begin{equation}
            \delta \geq \max \sum_{i}^{\mathcal{O}}\mathrm{Tr}(\mathcal{E}^\dagger(M_i)(\rho-\exp(\epsilon(d))\sigma)),
        \end{equation}
        where $\tau(\rho, \sigma)\leq d$. Notably, the highlighted advantage is that one only needs to focus on the properties of the matrix $\sum_i^{\mathcal{O}}\mathcal{E}^\dagger(M_i)$ without paying attention to how quantum states evolve in the circuit. Thus, the verification problem of QDP \footnote{Verifies that whether a quantum algorithm with a noisy circuit satisfies $(\epsilon,\delta)$-QDP or not.} is converted to a computational problem on a pre-prepared matrix~\cite{guan2023detecting}. This work illustrates that by calculating the maximum and minimum eigenvalues of this matrix, denoted as $\lambda_{\mathrm{min}}$ and $\lambda_{\mathrm{max}}$, respectively, the sufficient and necessary condition of $(\epsilon,\delta)$-QDP can be determined as
        \begin{equation}
        \label{eq:guan}
            \begin{cases}
                \epsilon \geq \ln((\underset{\mathcal{O}}{\max}\frac{\lambda_{\mathrm{max}}}{\lambda_{\mathrm{min}}}-1)d+1), \\
                \delta \geq \underset{\mathcal{O}}{\max}( d\lambda_{\mathrm{max}}-(\exp(\epsilon(d))+d-1)\lambda_{\mathrm{min}}).
            \end{cases}
        \end{equation}
        
        From~\equref{eq:guan}, the privacy budget $\epsilon$ is logarithmically related to $d$ and is an acceptable approach to regulate the privacy strength of the quantum algorithm by artificially adding noise to the input, which affects the distance between states.

    \subsection{Differential privacy preservation in measurements}
        \label{subsec:inherent_m}
        This part is dedicated to how DP and QDP are induced by internal or external randomization of quantum measurements. As introduced in Section~\ref{subsec:qa}, performing physical measurements on quantum hardware aims to estimate values for one qubit or distributions for multiple qubits, and more shots make it more accurate. However, limited by the computational cost and time, the number of shots is always finite, which leads to a statistical error, or so-called \textit{shot noise}.

        Like quantum noise in circuits, shot noise in measurements is likewise unavoidable. Physically, shot noise originates from a multitude of fluctuating discrete charges or light when running a real quantum device. Owing to statistical fluctuations, shot noise can be modeled by the \textit{central limit theorem}, i.e., Gaussian or normally distributed.
            
        Inspired by that, Li et al.~\cite{li2023differential} proposed the interesting idea of the utility of shot noise to establish DP-preserving quantum computing. For a quantum algorithm $\mathcal{A}$ with a noiseless circuit $\mathcal{E}$ and a set of projection measurements $\{M_i\}_{i\in \mathcal{O}}$, the measurement outcomes can be represented as $\mu_\rho=\mathrm{Tr}(M\mathcal{E}(\rho))$ and $\mu_\sigma=\mathrm{Tr}(M\mathcal{E}(\sigma))$, where a pair of states $\rho$ and $\sigma$ satisfies $\tau(\rho,\sigma)\leq d$. $\mathcal{A}=(\psi,\mathcal{E},\{M_i\}_{i\in \mathcal{O}})$ is said to satisfy $\epsilon$-QDP, which is only induced by shot noise with
        \begin{equation}
            \epsilon \leq \frac{{dr}}{(1 -{\mu}) {\mu}}\left[ \frac{9}{2}(1 - 2{\mu}) + \frac{3}{2}\sqrt{n_{\mathrm{shot}}}+ \frac{{dr} ({\mu} + {dr})n_{\mathrm{shot}}}{1 -{\mu}}\right],
        \end{equation}
        where $r$ is the largest rank of $M$, $\mu$ is $\min\{\mu_\rho,\mu_\sigma\}$, and $n_{\mathrm{shot}}$ is the number of measurement shots.
        
        For the quantum algorithm $\mathcal{A}$ with a noisy circuit $\mathcal{E}_\mathcal{N}$ modeled as a depolarizing channel and a set of projection measurements $\{M_i\}_{i\in \mathcal{O}}$, the outcomes of the measurements $\mu_\rho,\mu_\sigma$ are denoted the same as above. $\mathcal{A}=(\phi, \mathcal{E},\{M_i\}_{i\in \mathcal{O}})$ is said to satisfy $\epsilon$-QDP, which is induced by shot noise and depolarizing noise jointly with
        \begin{equation}
            {\epsilon} \leq \frac{\alpha}{1 -{\mu}}\left[ \frac{9}{2} (1 - 2{\mu}) + \frac{3}{2} \sqrt{n_{\mathrm{shot}}} +\frac{\alpha {\mu}^2 \left( 1 + \alpha \right)n_{\mathrm{shot}}}{1 -{\mu}} \right],
        \end{equation}
        where $\alpha=\frac{(1-p)dr}{p}2^n$, $n$ is the number of qubits and $p$ is the error probability of depolarization. The readers can learn the full proof from~\cite[Theorem 1 \& Theorem 2]{li2023differential}.

        Although QDP can be naturally induced by shot noise, DP-preserving quantum algorithms can also be implemented via artificial manipulation of quantum measurements. One approach involves perturbing the quantum measurement itself, whereas the other approach involves applying classical randomization to the measurement outcomes. Let us discuss both aspects.

       Angrisani and Kashefi~\cite{angrisani2022quantum} presented an example of a quantum RR mechanism that can map noiseless measurements to perturbed measurements. Assuming a set of POVM measurements $M=\{M_i\}_{i=1}^k$ that are mathematically modeled as positive semi-definite matrices and an intended privacy budget $\epsilon$, similar to the classic RR mechanism (\defref{def:rr}), we have perturbed measurements as 
        \begin{equation}
            M_i^{\mathrm{RR}}=\frac{\exp(\epsilon)-1}{\exp(\epsilon)-1+k}M_i+\frac{1}{\exp(\epsilon)-1+k}\mathbb{I}, \quad \mbox{where} \ i\in \mathcal{O}.
        \end{equation}
        Since perturbed measurements $\mathcal{M}_\mathrm{RR}(M)=\{M^{\mathrm{RR}}_i\}_{i\in\mathcal{O}}$ are $\epsilon$-LDP (\defref{def:ldp}), the quantum algorithm $\mathcal{A}=(\phi,\mathcal{E},\mathcal{M}_\mathrm{RR}(M))$ also satisfies $\epsilon$-LDP (\textit{cf.} \theref{theorem:postprocess}).

        On the other hand, the output of post-measurements is a discrete distribution, and it reasonably allows for privatization via classical randomization. 
        Aaronson and Rothblum~\cite{aronson2019gentle} claimed that a noisy measurement with a Laplace mechanism rather than a noisy circuit can provide both gentleness\footnote{The gentleness symbolizes the closeness between pre-measurement and post-measurement states~\cite[Definition 1]{aronson2019gentle}.} and privacy simultaneously for the quantum algorithm. If a measurement $\mathcal{M}_{\mathcal{N}}(M)$ is $\epsilon$-QDP on tensor product states, then $\mathcal{M}_{\mathcal{N}}(M)$ is $O(\epsilon\sqrt{n})$-gentle, where $n$ is the number of qubits.
        
        In~\cite{angrisani2023unifying}, $\epsilon$ and $\delta$ are given rigorously under the post-perturbed setting. Let $\rho,\sigma$ be a pair of neighboring states in which $\tau_{\mathrm{tr}}(\rho,\sigma)\leq d$ and $M$ are the POVM measurements. Let $\mathcal{M}_{\mathcal{N}}$ be an additive noise mechanism (\defref{Def:addnoise}) on the measurement outcomes. Owing to~\theref{theorem:ajc}, for all possible outcomes of $M$, such that $i,j$: $D_\alpha(\mathcal{M}_{\mathcal{N}}(i)\parallel\mathcal{M}_{\mathcal{N}}(j)\leq\delta$, the following inequality holds:
        \begin{equation}
            D_{\alpha}(\mathcal{M}_{\mathcal{N}}(M\rho)\parallel\mathcal{M}_{\mathcal{N}}(M\sigma)) \leq d\delta,
        \end{equation}
        where $\alpha=1+d(\exp(\epsilon)-1)$. 
        Thus, for POVM measurements $\mathcal{M}_{\mathcal{N}}$ and the quantum algorithm $\mathcal{A}=(\phi,\mathcal{E},\mathcal{M}_{\mathcal{N}}(M))$, if $\mathcal{N}\sim \mathrm{Lap}(0,\frac{|\mathcal{O}|}{\epsilon})$, $\mathcal{A}$ is said to satisfy $\log(1+d(\exp(\frac{|\mathcal{O}|}{\epsilon})-1))$-QDP; if $\mathcal{N}\sim\mathrm{Gau}(0,\frac{\sqrt{2\ln(1.25/\delta)}|\mathcal{O}|}{\epsilon})$, $\mathcal{A}$ is said to satisfy $\left(\log(1+d(\exp(\epsilon)-1)),d\delta\right)$-QDP. See more details in~\cite[Theorem 5.3]{angrisani2023unifying}.

\section{Challenges and future work directions}
\label{sec:challenges}
Despite some pioneering work on DP-preserving quantum algorithms, related research is still in its infancy and requires more in-depth study. In this section, we present some existing challenges and potential future directions of DP in quantum computing.

\subsection{Unified benchmark for DP-preserving quantum algorithms}
    There is currently no standardized privacy metric or unified benchmark to evaluate these DP-preserving quantum algorithms with different distance metrics (\tabref{tab:dis_metric}) and different randomizations in some phases. How to quantify the privacy parameters of these QDP formulations is a concern. A unified benchmark for QDP would help us compare various approaches and arrange optimal DP schemes that balance performance and privacy in quantum algorithms.

\subsection{Built-in QDP implementation for quantum simulators}
    Realistic quantum devices are uncommon and costly, so software or cloud service providers release software-level quantum simulators~\cite{aleksandrowicz2019qiskit, bergholm2018pennylane, broughton2020tensorflow, wang2022quantumnas, kim2023cuda} or cloud access to quantum computers~\cite{amazon2020braket, ms2023azure} for the convenience of researchers and developers. However, while the privacy issue arises in the quantum scenario, neither has realized a built-in privacy module, e.g., the implementation of QDP. Akin to TensorFlow Privacy (TF Privacy)~\cite{google2021tfprivacy}, we believe that it is necessary to equip future quantum simulation software or cloud platforms with built-in QDP implementation.

\subsection{Privacy audit of QDP mechanisms}
    In the classical scenario, researchers struggle to find a tighter lower bound of privacy parameters~\cite{jagielski2020auditing,nasr2023tight,steinke2023privacy} and detect violations when the lower and upper bounds are contradictory~\cite{bichsel2018dp,ding2018detecting,tramer2022debugging}. The upper and lower privacy bounds for QDP are also of concern. The DP-preserving quantum algorithms introduced in Section~\ref{Sec:DP_QA} theoretically provide a \textit{upper bound} of the privacy parameters $\epsilon$ and $\delta$ (shown in~\tabref{tab:qdp_qc}), whereas Guan's work~\cite{guan2023detecting} first presented a lower bound of privacy parameters, as shown in \equref{eq:guan}. Nevertheless, the next step is to experimentally implement QDP mechanisms and privacy auditing in the quantum scenario to empirically obtain tighter lower bounds for these DP-preserving quantum algorithms.
% \subsection{Privacy amplification of QDP} 

\subsection{Privacy account for QDP}
    One of the most important properties of DP is composability, which dictates how to combine multiple privacy mechanisms~\cite{dwork2010boosting,kairouz2015composition}. Furthermore, the privacy accountant, which is responsible for keeping track of the privacy loss to maintain a moderate privacy cost, possesses a tighter composition bound and is widely employed in the privacy-preserving deep learning model~\cite{abadi2016deep}. The composition theorem for QDP was also established~\cite{zhou2017differential,hirche2023quantum}, as shown in~\theref{theorem:composition}. However, there is still a lack of a privacy account for QDP, which has potential utility in future privacy-preserving quantum algorithms.

\section{Conclusion}
\label{sec:conclusion}
    As practical quantum algorithms are coming into use, it is reasonable to proactively consider privacy concerns for any possible eventualities. In this survey, we report selective papers regarding QDP, which is a promising approach toward privacy-preserving quantum computing and has generated a growing interest for institute researchers. We provide a categorization of DP-preserving quantum algorithms here to offer better intuitions rather than a definitive/authoritative guide. The overview shows that any phase of a raw quantum algorithm can yield appropriate privacy via internal or external randomization. Therefore, privacy can be tuned regularly in quantum algorithms by not only natural quantum noise but also artificial noise addition. We hope that the work we have done will serve as valuable inspiration for future researchers.

\balance
\bibliographystyle{IEEEtran}
\bibliography{reference}

\begin{thebibliography}{10}
\providecommand{\url}[1]{#1}
\csname url@samestyle\endcsname
\providecommand{\newblock}{\relax}
\providecommand{\bibinfo}[2]{#2}
\providecommand{\BIBentrySTDinterwordspacing}{\spaceskip=0pt\relax}
\providecommand{\BIBentryALTinterwordstretchfactor}{4}
\providecommand{\BIBentryALTinterwordspacing}{\spaceskip=\fontdimen2\font plus
\BIBentryALTinterwordstretchfactor\fontdimen3\font minus \fontdimen4\font\relax}
\providecommand{\BIBforeignlanguage}[2]{{%
\expandafter\ifx\csname l@#1\endcsname\relax
\typeout{** WARNING: IEEEtran.bst: No hyphenation pattern has been}%
\typeout{** loaded for the language `#1'. Using the pattern for}%
\typeout{** the default language instead.}%
\else
\language=\csname l@#1\endcsname
\fi
#2}}
\providecommand{\BIBdecl}{\relax}
\BIBdecl

\bibitem{grover1996fast}
L.~K. Grover, ``A fast quantum mechanical algorithm for database search,'' in \emph{Proceedings of Annual ACM SIGACT Symposium on Theory of Computing (STOC'96)}, July 1996, pp. 212--219.

\bibitem{shor1994algorithms}
P.~W. Shor, ``Algorithms for quantum computation: discrete logarithms and factoring,'' in \emph{Proceedings of Annual Symposium on Foundations of Computer Science (FOCS'94)}, November 1994, pp. 124--134.

\bibitem{harrow2009quantum}
A.~W. Harrow, A.~Hassidim, and S.~Lloyd, ``Quantum algorithm for linear systems of equations,'' \emph{Physical Review Letters}, vol. 103, no.~15, p. 150502, October 2009.

\bibitem{zhong2020quantum}
H.-S. Zhong, H.~Wang, Y.-H. Deng, M.-C. Chen, L.-C. Peng, Y.-H. Luo, J.~Qin, D.~Wu, X.~Ding, Y.~Hu \emph{et~al.}, ``Quantum computational advantage using photons,'' \emph{Science}, vol. 370, no. 6523, pp. 1460--1463, December 2020.

\bibitem{arute2019quantum}
F.~Arute, K.~Arya, R.~Babbush, D.~Bacon, J.~C. Bardin, R.~Barends, R.~Biswas, S.~Boixo, F.~G. Brandao, D.~A. Buell \emph{et~al.}, ``Quantum supremacy using a programmable superconducting processor,'' \emph{Nature}, vol. 574, no. 7779, pp. 505--510, October 2019.

\bibitem{cerezo2021variational}
M.~Cerezo, A.~Arrasmith, R.~Babbush, S.~C. Benjamin, S.~Endo, K.~Fujii, J.~R. McClean, K.~Mitarai, X.~Yuan, L.~Cincio \emph{et~al.}, ``Variational quantum algorithms,'' \emph{Nature Reviews Physics}, vol.~3, no.~9, pp. 625--644, August 2021.

\bibitem{peruzzo2014variational}
A.~Peruzzo, J.~McClean, P.~Shadbolt, M.-H. Yung, X.-Q. Zhou, P.~J. Love, A.~Aspuru-Guzik, and J.~L. O’brien, ``A variational eigenvalue solver on a photonic quantum processor,'' \emph{Nature Communications}, vol.~5, no.~1, p. 4213, July 2014.

\bibitem{farhi2014quantum}
E.~Farhi, J.~Goldstone, and S.~Gutmann, ``A quantum approximate optimization algorithm,'' \emph{arXiv preprint arXiv:1411.4028}, November 2014.

\bibitem{rebentrost2014quantum}
P.~Rebentrost, M.~Mohseni, and S.~Lloyd, ``Quantum support vector machine for big data classification,'' \emph{Physical Review Letters}, vol. 113, no.~13, p. 130503, September 2014.

\bibitem{lloyd2018quantum}
S.~Lloyd and C.~Weedbrook, ``Quantum generative adversarial learning,'' \emph{Physical Review Letters}, vol. 121, no.~4, p. 040502, April 2018.

\bibitem{cong2019quantum}
I.~Cong, S.~Choi, and M.~D. Lukin, ``Quantum convolutional neural networks,'' \emph{Nature Physics}, vol.~15, no.~12, pp. 1273--1278, August 2019.

\bibitem{bausch2020recurrent}
J.~Bausch, ``Recurrent quantum neural networks,'' \emph{Advances in Neural Information Processing Systems (NeurIPS'20)}, pp. 1368--1379, December 2020.

\bibitem{cao2018potential}
Y.~Cao, J.~Romero, and A.~Aspuru-Guzik, ``Potential of quantum computing for drug discovery,'' \emph{IBM Journal of Research and Development}, vol.~62, no.~6, pp. 6--1, November 2018.

\bibitem{de2021materials}
N.~P. De~Leon, K.~M. Itoh, D.~Kim, K.~K. Mehta, T.~E. Northup, H.~Paik, B.~Palmer, N.~Samarth, S.~Sangtawesin, and D.~W. Steuerman, ``Materials challenges and opportunities for quantum computing hardware,'' \emph{Science}, vol. 372, no. 6539, p. 2823, April 2021.

\bibitem{orus2019quantum}
R.~Or{\'u}s, S.~Mugel, and E.~Lizaso, ``Quantum computing for finance: Overview and prospects,'' \emph{Reviews in Physics}, vol.~4, p. 100028, November 2019.

\bibitem{herman2023quantum}
D.~Herman, C.~Googin, X.~Liu, Y.~Sun, A.~Galda, I.~Safro, M.~Pistoia, and Y.~Alexeev, ``Quantum computing for finance,'' \emph{Nature Reviews Physics}, vol.~5, no.~8, pp. 450--465, July 2023.

\bibitem{berger2021quantum}
C.~Berger, A.~Di~Paolo, T.~Forrest, S.~Hadfield, N.~Sawaya, M.~St{\k{e}}ch{\l}y, and K.~Thibault, ``Quantum technologies for climate change: Preliminary assessment,'' \emph{arXiv preprint arXiv:2107.05362}, July 2021.

\bibitem{xu2023classification}
C.~Xu, F.~Erata, and J.~Szefer, ``Classification of quantum computer fault injection attacks,'' \emph{arXiv preprint arXiv:2309.05478}, September 2023.

\bibitem{chu2023qtrojan}
C.~Chu, L.~Jiang, M.~Swany, and F.~Chen, ``Qtrojan: A circuit backdoor against quantum neural networks,'' in \emph{IEEE International Conference on Acoustics, Speech and Signal Processing (ICASSP'23)}, June 2023, pp. 1--5.

\bibitem{chu2023qdoor}
C.~Chu, F.~Chen, P.~Richerme, and L.~Jiang, ``Qdoor: Exploiting approximate synthesis for backdoor attacks in quantum neural networks,'' in \emph{IEEE International Conference on Quantum Computing and Engineering (QCE'23)}, September 2023, pp. 1098--1106.

\bibitem{amazon2020braket}
\BIBentryALTinterwordspacing
{Amazon Web Services}. (2020) Amazon braket. [Online]. Available: \url{https://aws.amazon.com/braket}
\BIBentrySTDinterwordspacing

\bibitem{ms2023azure}
\BIBentryALTinterwordspacing
{Microsoft Azure}. (2023) Azure quantum cloud service. [Online]. Available: \url{https://azure.microsoft.com/en-us/products/quantum}
\BIBentrySTDinterwordspacing

\bibitem{dwork2014algorithmic}
C.~Dwork, A.~Roth \emph{et~al.}, ``The algorithmic foundations of differential privacy,'' \emph{Foundations and Trends{\textregistered} in Theoretical Computer Science}, vol.~9, no. 3--4, pp. 211--407, August 2014.

\bibitem{dwork2006calibrating}
C.~Dwork, F.~McSherry, K.~Nissim, and A.~Smith, ``Calibrating noise to sensitivity in private data analysis,'' in \emph{Theory of Cryptography Conference (TCC'06)}, March 2006, pp. 265--284.

\bibitem{abadi2016deep}
M.~Abadi, A.~Chu, I.~Goodfellow, H.~B. McMahan, I.~Mironov, K.~Talwar, and L.~Zhang, ``Deep learning with differential privacy,'' in \emph{ACM SIGSAC Conference on Computer and Communications Security (CCS'16)}, October 2016, pp. 308--318.

\bibitem{zhou2017differential}
L.~Zhou and M.~Ying, ``Differential privacy in quantum computation,'' in \emph{IEEE Computer Security Foundations Symposium (CSF'17)}, August 2017, pp. 249--262.

\bibitem{du2021quantum}
Y.~Du, M.-H. Hsieh, T.~Liu, D.~Tao, and N.~Liu, ``Quantum noise protects quantum classifiers against adversaries,'' \emph{Physical Review Research}, vol.~3, no.~2, p. 023153, May 2021.

\bibitem{nielsen2010quantum}
M.~A. Nielsen and I.~L. Chuang, \emph{Quantum computation and quantum information}.\hskip 1em plus 0.5em minus 0.4em\relax Cambridge University Press, December 2010.

\bibitem{schuld2021supervised}
M.~Schuld, ``Supervised quantum machine learning models are kernel methods,'' \emph{arXiv preprint arXiv:2101.11020}, January 2021.

\bibitem{dwork2006differential}
C.~Dwork, ``Differential privacy,'' in \emph{International Colloquium on Automata, Languages, and Programming (ICALP'06)}, July 2006, pp. 1--12.

\bibitem{dwork2010boosting}
C.~Dwork, G.~N. Rothblum, and S.~Vadhan, ``Boosting and differential privacy,'' in \emph{IEEE Symposium on Foundations of Computer Science (FOCS'10)}, October 2010, pp. 51--60.

\bibitem{erlingsson2014rappor}
{\'U}.~Erlingsson, V.~Pihur, and A.~Korolova, ``Rappor: Randomized aggregatable privacy-preserving ordinal response,'' in \emph{ACM SIGSAC Conference on Computer and Communications Security (CCS'14)}, November 2014, pp. 1054--1067.

\bibitem{gong2022enhancing}
W.~Gong, D.~Yuan, W.~Li, and D.-L. Deng, ``Enhancing quantum adversarial robustness by randomized encodings,'' \emph{Physical Review Research}, vol.~6, no.~2, p. 023020, April 2024.

\bibitem{hirche2023quantum}
C.~Hirche, C.~Rouz{\'e}, and D.~S. Fran{\c{c}}a, ``Quantum differential privacy: An information theory perspective,'' \emph{IEEE Transactions on Information Theory (TIT)}, vol.~69, no.~9, pp. 5771--5787, September 2023.

\bibitem{aronson2019gentle}
S.~Aaronson and G.~N. Rothblum, ``Gentle measurement of quantum states and differential privacy,'' in \emph{ACM SIGACT Symposium on Theory of Computing (STOC'19)}, June 2019, pp. 322--333.

\bibitem{angrisani2023unifying}
M.~Angrisani, Armando, E.~Doosti, and Kashefi, ``A unifying framework for differentially private quantum algorithms,'' \emph{arXiv preprint arXiv:2307.04733}, July 2023.

\bibitem{verdon2019learning}
G.~Verdon, M.~Broughton, J.~R. McClean, K.~J. Sung, R.~Babbush, Z.~Jiang, H.~Neven, and M.~Mohseni, ``Learning to learn with quantum neural networks via classical neural networks,'' \emph{arXiv preprint arXiv:1907.05415}, July 2019.

\bibitem{de2021quantum}
G.~De~Palma, M.~Marvian, D.~Trevisan, and S.~Lloyd, ``The quantum wasserstein distance of order 1,'' \emph{IEEE Transactions on Information Theory (TIT)}, vol.~67, no.~10, pp. 6627--6643, May 2021.

\bibitem{nuradha2023quantum}
T.~Nuradha, Z.~Goldfeld, and M.~M. Wilde, ``Quantum pufferfish privacy: A flexible privacy framework for quantum systems,'' \emph{IEEE Transactions on Information Theory (TIT)}, vol.~70, no.~8, pp. 5731--5762, May 2024.

\bibitem{senekane2017privacy}
M.~Senekane, M.~Mafu, and B.~M. Taele, ``Privacy-preserving quantum machine learning using differential privacy,'' in \emph{Proceedings of IEEE AFRICON}, September 2017, pp. 1432--1435.

\bibitem{yoshida2020classical}
Y.~Yoshida and M.~Hayashi, ``Classical mechanism is optimal in classical-quantum differentially private mechanisms,'' in \emph{IEEE International Symposium on Information Theory (ISIT'20)}, June 2020, pp. 1973--1977.

\bibitem{du2022quantum}
Y.~Du, M.-H. Hsieh, T.~Liu, S.~You, and D.~Tao, ``Quantum differentially private sparse regression learning,'' \emph{IEEE Transactions on Information Theory (TIT)}, vol.~68, no.~8, pp. 5217--5233, April 2022.

\bibitem{angrisani2022differential}
A.~Angrisani, M.~Doosti, and E.~Kashefi, ``Differential privacy amplification in quantum and quantum-inspired algorithms,'' in \emph{International Conference on Learning Representations Workshop (ICLR Workshop'22)}, April 2022.

\bibitem{guan2023detecting}
J.~Guan, W.~Fang, M.~Huang, and M.~Ying, ``Detecting violations of differential privacy for quantum algorithms,'' in \emph{ACM SIGSAC Conference on Computer and Communications Security (CCS'23)}, November 2023.

\bibitem{li2021quantum}
W.~Li, S.~Lu, and D.-L. Deng, ``Quantum federated learning through blind quantum computing,'' \emph{Science China Physics, Mechanics \& Astronomy}, vol.~64, no.~10, p. 100312, September 2021.

\bibitem{watkins2023quantum}
W.~M. Watkins, S.~Y.-C. Chen, and S.~Yoo, ``Quantum machine learning with differential privacy,'' \emph{Scientific Reports}, vol.~13, no.~1, p. 2453, February 2023.

\bibitem{rofougaran2023federated}
R.~Rofougaran, S.~Yoo, H.-H. Tseng, and S.~Y.-C. Chen, ``Federated quantum machine learning with differential privacy,'' in \emph{IEEE International Conference on Acoustics, Speech and Signal Processing (ICASSP'24)}, April 2024, pp. 9811--9815.

\bibitem{li2023differential}
Y.~Li, Y.~Zhao, X.~Zhang, H.~Zhong, M.~Pan, and C.~Zhang, ``Differential privacy preserving quantum computing via projection operator measurements,'' in \emph{International Conference on Quantum Communications, Networking, and Computing (QCNC'24)}, July 2024.

\bibitem{angrisani2022quantum}
A.~Angrisani and E.~Kashefi, ``Quantum statistical query model and local differential privacy,'' in \emph{International Conference on Machine Learning Workshop (ICML Workshop'21)}, July 2021.

\bibitem{dankert2009exact}
C.~Dankert, R.~Cleve, J.~Emerson, and E.~Livine, ``Exact and approximate unitary 2-designs and their application to fidelity estimation,'' \emph{Physical Review A}, vol.~80, no.~1, p. 012304, July 2009.

\bibitem{vovrosh2021simple}
J.~Vovrosh, K.~E. Khosla, S.~Greenaway, C.~Self, M.~S. Kim, and J.~Knolle, ``Simple mitigation of global depolarizing errors in quantum simulations,'' \emph{Physical Review E}, vol. 104, no.~3, p. 035309, September 2021.

\bibitem{huang2023certified}
J.-C. Huang, Y.-L. Tsai, C.-H.~H. Yang, C.-F. Su, C.-M. Yu, P.-Y. Chen, and S.-Y. Kuo, ``Certified robustness of quantum classifiers against adversarial examples through quantum noise,'' in \emph{IEEE International Conference on Acoustics, Speech and Signal Processing (ICASSP'23)}, June 2023, pp. 1--5.

\bibitem{aleksandrowicz2019qiskit}
G.~Aleksandrowicz, T.~Alexander, P.~Barkoutsos, L.~Bello, Y.~Ben-Haim, D.~Bucher, F.~J. Cabrera-Hern{\'a}ndez, J.~Carballo-Franquis, A.~Chen, C.-F. Chen \emph{et~al.}, ``Qiskit: An open-source framework for quantum computing,'' \emph{Accessed on: March}, vol.~16, March 2019.

\bibitem{bergholm2018pennylane}
V.~Bergholm, J.~Izaac, M.~Schuld, C.~Gogolin, S.~Ahmed, V.~Ajith, M.~S. Alam, G.~Alonso-Linaje, B.~AkashNarayanan, A.~Asadi \emph{et~al.}, ``Pennylane: Automatic differentiation of hybrid quantum-classical computations,'' \emph{arXiv preprint arXiv:1811.04968}, November 2018.

\bibitem{broughton2020tensorflow}
M.~Broughton, G.~Verdon, T.~McCourt, A.~J. Martinez, J.~H. Yoo, S.~V. Isakov, P.~Massey, R.~Halavati, M.~Y. Niu, A.~Zlokapa \emph{et~al.}, ``Tensorflow quantum: A software framework for quantum machine learning,'' \emph{arXiv preprint arXiv:2003.02989}, March 2020.

\bibitem{wang2022quantumnas}
H.~Wang, Y.~Ding, J.~Gu, Y.~Lin, D.~Z. Pan, F.~T. Chong, and S.~Han, ``Quantumnas: Noise-adaptive search for robust quantum circuits,'' in \emph{IEEE International Symposium on High-Performance Computer Architecture (HPCA'22)}, April 2022, pp. 692--708.

\bibitem{kim2023cuda}
J.-S. Kim, A.~McCaskey, B.~Heim, M.~Modani, S.~Stanwyck, and T.~Costa, ``Cuda quantum: The platform for integrated quantum-classical computing,'' in \emph{ACM/IEEE Design Automation Conference (DAC'23)}, July 2023, pp. 1--4.

\bibitem{google2021tfprivacy}
\BIBentryALTinterwordspacing
{Google TensorFlow}. (2021) Tensorflow privacy. [Online]. Available: \url{https://www.tensorflow.org/responsible_ai/privacy}
\BIBentrySTDinterwordspacing

\bibitem{jagielski2020auditing}
M.~Jagielski, J.~Ullman, and A.~Oprea, ``Auditing differentially private machine learning: How private is private sgd?'' in \emph{Advances in Neural Information Processing Systems (NeurIPS'20)}, vol.~33, December 2020, pp. 22\,205--22\,216.

\bibitem{nasr2023tight}
M.~Nasr, J.~Hayes, T.~Steinke, B.~Balle, F.~Tram{\`e}r, M.~Jagielski, N.~Carlini, and A.~Terzis, ``Tight auditing of differentially private machine learning,'' in \emph{USENIX Security Symposium (USENIX Security’23)}, August 2023.

\bibitem{steinke2023privacy}
T.~Steinke, M.~Nasr, and M.~Jagielski, ``Privacy auditing with one (1) training run,'' in \emph{Advances in Neural Information Processing Systems (NeurIPS'23)}, December 2023.

\bibitem{bichsel2018dp}
B.~Bichsel, T.~Gehr, D.~Drachsler-Cohen, P.~Tsankov, and M.~Vechev, ``Dp-finder: Finding differential privacy violations by sampling and optimization,'' in \emph{ACM SIGSAC Conference on Computer and Communications Security (CCS'18)}, October 2018, pp. 508--524.

\bibitem{ding2018detecting}
Z.~Ding, Y.~Wang, G.~Wang, D.~Zhang, and D.~Kifer, ``Detecting violations of differential privacy,'' in \emph{ACM SIGSAC Conference on Computer and Communications Security (CCS'18)}, October 2018, pp. 475--489.

\bibitem{tramer2022debugging}
F.~Tramer, A.~Terzis, T.~Steinke, S.~Song, M.~Jagielski, and N.~Carlini, ``Debugging differential privacy: A case study for privacy auditing,'' \emph{arXiv preprint arXiv:2202.12219}, February 2022.

\bibitem{kairouz2015composition}
P.~Kairouz, S.~Oh, and P.~Viswanath, ``The composition theorem for differential privacy,'' in \emph{International Conference on Machine Learning (ICML'15)}, June 2015, pp. 1376--1385.

\end{thebibliography}

\end{document}